\documentclass[hdvipdfmx]{ptephy_v1}

\usepackage{subfig} 

\usepackage{graphics}

\usepackage{dcolumn}

\newcommand{\Kslash}{K \hspace{-8pt}/ }
\newcommand{\delslash}{\partial \hspace{-6pt}/ }

\usepackage[normalem]{ulem} 
\usepackage{color}

\renewcommand\sout{\bgroup \color{red} \ULdepth=-.5ex \ULset}

\begin{document}

\title{Structure of $\eta^{\prime}$ mesonic nuclei 
in a relativistic mean field theory}

\author[1]{Daisuke Jido\thanks{Present address: Department of Physics, Tokyo Institute of Technology, Meguro, Tokyo 152-8551, Japan}}
\author[1]{Hanayo Masutani}
\author[2]{Satoru Hirenzaki}

\affil[1]{Department of Physics, Tokyo Metropolitan University, 
1-1 Minami-Osawa, 
Hachioji, Tokyo 192-0397, Japan}
\affil[2]{Department of Physics, Nara Women's University, Nara 630-8506, Japan\email{jido@th.phys.titech.ac.jp}}

\begin{abstract}%
The structure and the energy spectrum of the $\eta^{\prime}$ mesonic nuclei are investigated 
in a relativistic mean field theory. 
One expects a substantial
attraction for the $\eta^{\prime}$ meson in finite nuclei  
due to the partial restoration of chiral symmetry in the nuclear medium.
Such a hadronic scale interaction  for the $\eta^{\prime}$ mesonic
nuclei may provide modification of the nuclear structure. 
The relativistic mean field theory is a self-contained model for finite nuclei 
which provides the saturation property within the model, and is good to 
investigate the structure change of the nucleus induced by the $\eta^{\prime}$ meson. 
Using the local density approximation for the mean fields, we solve the equations of motion 
for the nucleons and the $\eta^{\prime}$ meson self-consistently, and obtain 
the nuclear density distribution and the $\eta^{\prime}$ energy spectrum 
for the $\eta^{\prime}$ mesonic nuclei. We take $^{12}$C, $^{16}$O and $^{40}$Ca for the target nuclei. 
We find several bound states of the $\eta^{\prime}$ meson for these nuclei thanks to the attraction 
for $\eta^{\prime}$ in nuclei. We also find a sufficient change of the nuclear structure especially for the $1s$ bound state of $\eta^{\prime}$. 
This implies that the production of the $1s$ bound state in nuclear reaction 
may be suppressed.
\end{abstract}

\subjectindex{xxxx, xxx}

\maketitle

\section{Introduction}

Partial restoration of chiral symmetry, which is incomplete restoration 
of spontaneous breaking of chiral symmetry with a sufficient reduction of
chiral condensate, has been suggested  
in deeply bound pionic atoms~\cite{Suzuki:2002ae} and 
low-energy pion nucleus elastic scattering~\cite{Friedman:2004jh}
with a help of theoretical considerations~\cite{Kolomeitsev:2002gc,Jido:2008bk}.
Thanks to this phenomenological finding, it can be believed that 
partial restoration of chiral symmetry really takes place in the nuclear medium.
Based on linear density approximations, it has been found that 
the magnitude of the quark condensate is reduced by 
about 30\%~\cite{Suzuki:2002ae,Jido:2008bk}. The next steps along this line
are investigation of the density dependence of the quark condensate beyond the linear 
density approximation as done in, for instance, Refs.~\cite{Fritsch:2004nx,Goda:2013bka,Goda:2013npa}
and systematic studies of partial restoration of chiral symmetry in other physical systems. 
One of the latter examples is the in-medium properties of the $\eta^{\prime}$ meson. 
It has been pointed out in 
a study of the QCD correlation function~\cite{Lee:1996zy} and
in a group theoretical argument~\cite{Jido:2011pq,Jido:2012av} that 
in order to affect the U$_{A}$(1) anomaly on the $\eta^{\prime}$ mass the SU(3) chiral symmetry 
is necessarily broken and, thus, the $\eta^{\prime}$ mass should be reduced in the nuclear matter
where partial restoration of chiral symmetry takes place. So far, many works were devoted to
the in-medium calculation of the $\eta^{\prime}$ meson~\cite{Pisarski:1983ms,Kapusta:1995ww,Bass:2005hn,Bernard:1987sg,Tsushima:1998qp,Nagahiro:2011fi}. 
Along the context of the partial restoration of chiral symmetry in the nuclear medium, 
the reduction of the $\eta^{\prime}$ mass at the saturation density was estimated 
as about 150 MeV in the NJL model~\cite{Costa:2002gk,Nagahiro:2006} and 
as about 80 MeV in the linear $\sigma$ mode~\cite{Sakai:2013nba,Sakai:2016vcl}. 
There are also theoretical studies suggesting that the $\eta^{\prime}$ mass is insensitive to the nuclear density, such as Refs.~\cite{Bernard:1987sx,Fejos:2017kpq,Fejos:2018dyy}.
Thus, the in-medium properties of the $\eta^{\prime}$ meson are interesting to be 
studied also as one of the examples of the phenomena under partial restoration of 
chiral symmetry in the nuclear matter. 

The reduction of the mass in the infinite nuclear matter 
is described by an attractive in-medium scalar self-energy of the hadron.
The self-energy in medium has density dependence.  
In a finite nucleus, because the density depends on the position in the nucleus, the attractive self-energy serves a position-dependent attractive 
potential with the local density approximation.
In particular, the mass reduction induced by partial restoration of chiral symmetry is described by the interaction with the $\sigma$ field. 
It is natural to have also
repulsive interactions for the in-medium mesons as medium effects, but in the case of the $\eta^{\prime}$ meson it is known that the Weinberg-Tomozawa interaction vanishes for the $\eta^{\prime}N$ channel and it may be expected to have small repulsion from the $\omega$ meson exchange, which is a significant source of the repulsive interaction for the nuclear force.
Thus, the possible mass reduction of the $\eta^{\prime}$ meson 
in the nuclear brings about the corresponding attractive scalar
potential to the $\eta^{\prime}$ meson 
in nuclei, and one expects some bound states of the $\eta^{\prime}$ meson in nuclei. The formation 
spectrum of the $\eta^{\prime}$ bound state in a nucleus was first calculated in Ref.~\cite{Nagahiro:2004qz}
for a $(\gamma,p)$ reaction with a nuclear target. Later a comprehensive study for the $\eta^{\prime}$ 
bound state formation spectrum in a $(p,d)$ reaction was done in Ref.~\cite{Nagahiro:2012aq}
and an experimental feasibility to observe the $\eta^{\prime}$ bound states in nuclei 
was investigated in Ref.~\cite{Itahashi:2012ut}. Experimental measurement of 
the $\eta^{\prime}$ bound states in $^{11}$C has been done in GSI using the $^{12}$C$(p,d)$ 
reaction~\cite{Tanaka:2016bcp,Tanaka:2017cme}.
Unfortunately a clear peak structure was not observed in the experiment. Still one has to make an 
effort to reduce background against the signal of the $\eta^{\prime}$ bound state. 
The information of the optical potential for the $\eta^{\prime}$ meson in nuclei was deduced from the $\eta^{\prime}$ photoproduction experiments on nuclear targets in Refs.~\cite{Nanova:2013fxl,Nanova:2016cyn} for the real part and in Refs.~\cite{Nanova:2012vw,Friedrich:2016cms} for the imaginary part. The scattering length of the $\eta^{\prime}p$ interaction was also extracted from the $pp \to pp\eta^{\prime}$ reaction~\cite{Czerwinski:2014yot}. A formation experiment of the $\eta^{\prime}$ mesonic nuclei in the $^{12}$C$(\gamma,p)$ reaction by the LEPS2 collaboration is going on \cite{Muramatsu:2013tdv}.

In this paper we investigate the $\eta^{\prime}$ bound states in nuclei using 
a relativistic mean field theory,
in which we regard the $\eta^{\prime}$ meson as a constituent of the nucleus and introduce attractive interaction between the $\eta^{\prime}$ meson and the $\sigma$ mean field motivated by the $\eta^{\prime}$ mass reduction under partial restoration of chiral symmetry in the nuclear medium.
In the previous calculations of the $\eta^{\prime}$ bound states~\cite{Jido:2011pq,Jido:2012av},
the nuclear density was assumed 
as a normal nucleus and the $\eta^{\prime}$ bound states were calculated with a 
fixed potential. It is natural that, if the interaction between the $\eta^{\prime}$ and nucleus
is strong enough for producing the 100 MeV mass reduction, the nuclear structure 
could be changed due to the strong interaction. 
In the relativistic mean field theory,
the saturation property is reproduced within the model and the nuclear structure 
is obtained by solving Dirac equation for the nucleon under the presence of the $\sigma$, 
$\omega$, $\rho$ and electric mean fields. 
Thanks to automatic implementation of the spin-orbit force with correct strength 
in the relativistic formulation, the magic number of the nuclear structure is successfully reproduced.
The relativistic nuclear field theory was introduced by Refs.~\cite{Johnson:1955zz,Duerr:1956zz} and was developed in Refs.~\cite{Walecka:1974qa,Boguta:1977xi}. References~\cite{Boguta:1977fu,Serr:1978kh,Serot:1979cc} applied for finite nuclei and investigated the properties of nucleus, such as the nuclear density distribution.
The relativistic mean field theory was applied for the investigation of kaonic nuclei in 
Refs.~\cite{Gazda:2007wd,Gazda:2008xd,Gazda:2012}.
In the relativistic mean field theory, the nuclear matter and nuclei are reproduced within the model,
one can calculate the modification of the nuclear matter and nuclei under the presence 
of the $\eta^{\prime}$ meson as an impurity. Such back-reaction is very important 
for the investigation of the structure of the $\eta^{\prime}$-nucleus bound system. 
In this paper, we consider $^{12}$C, $^{16}$O and $^{40}$Ca as target nuclei and 
show the $\eta^{\prime}$ bound state spectra and nuclear density profiles in the presence 
of the $\eta^{\prime}$ meson in the nuclei. 
We assume that nuclear absorption of the $\eta^{\prime}$ meson in the mesonic nucleus is not considered, and thus the potential for the $\eta^{\prime}$ meson is to be pure real. 
One would expect that heavy nuclei could be good for the observation of the $\eta^{\prime}$ 
bound states in a nucleus. Indeed, the more bound states can be formed in the heavier nuclei.
Nevertheless, the formation spectra of the $\eta^{\prime}$ mesonic nuclei can be complicated,
because many peaks from these bound states are overlapped~\cite{Nagahiro:2012aq}. 
It may be hard to identify the $\eta^{\prime}$ bound states from such complicated spectra.

\section{Formulation}

We investigate bound systems of an $\eta^{\prime}$ meson in a nucleus in the relativistic mean field theory. 
The $\eta^{\prime}$ meson and the nucleons are constituents of the bound systems. Thus, we treat 
the $\eta^{\prime}$ meson as a matter field as well as the nucleon. 
The interactions between the constituents, or the $NN$ and $\eta^{\prime}N$ interactions, 
are mediated by the bosons. Here we introduce the $\sigma$, $\omega$ and $\rho$ mesons for 
the strong interactions and photon for the electric interaction between protons. 

The Lagrangian that we use in the present work is given by
\begin{eqnarray}
\mathcal{L}&=&\bar{\psi}\bigl[i\gamma_\mu\bigl\{\partial^\mu+ig_\omega\omega^\mu
+ig_\rho\rho^\mu\frac{\tau^3}{2}
+ie A^{\mu} \frac{1+\tau^{3}}{2} \bigr\}\bigr]\psi
-\bar{\psi}\bigl(m-g_\sigma\sigma\bigr)\psi\nonumber\\
&&+\frac{1}{2}\partial_\mu\sigma\partial^\mu\sigma-\frac{1}{2}m_\sigma^2\sigma^2-\frac{1}{3}bmg_\sigma^3\sigma^3-\frac{1}{4}cg_\sigma^4\sigma^4\nonumber\\
&&-\frac{1}{4}\omega_{\mu\nu}\omega^{\mu\nu}+\frac{1}{2}m_\omega^2\omega^2-\frac{1}{4}R_{\mu\nu}R^{\mu\nu}+\frac{1}{2}m_\rho^2\rho_{03}^2 
- \frac14 F_{\mu\nu} F^{\mu\nu}\nonumber\\
&&+ \frac12 \partial_{\mu} \eta^{\prime} \partial^{\mu} \eta^{\prime} - \frac{1}{2} m_{\eta^{\prime}}^{2} \eta^{\prime2}
+ g_{\sigma \eta^{\prime}} m_{\eta^{\prime}} \eta^{\prime2} \sigma.
  \label{eq:Lag}
\end{eqnarray}
Here $\psi$ is the nucleon iso-doublet field given as
\begin{equation}
   \psi  = \left(\begin{array}{c} p \\ n \end{array} \right),
\end{equation}
and $\tau^{3}$ is the third component of the Pauli matrix for the isospin space. 
We assume the isospin symmetry and $m$ is the nucleon mass. The nucleon 
couples to the meson fields $\omega^{\mu}$, $\rho^{\mu}$ and $\sigma$,
with the coupling constants $g_{\omega}$, $g_{\rho}$ and $g_{\sigma}$, respectively,
and to the photon field $A^{\mu}$ with electric charge $e$. 
Since we consider the nuclear matter ground state, or the lowest-energy configuration of the nucleons, we take into account only the neutral $\rho$ field.
The field tensors for the vector fields are given by
\begin{equation}
  \omega_{\mu\nu} = \partial_{\mu} \omega_{\nu} - \partial_{\nu} \omega_{\mu}, \qquad
  R_{\mu\nu} = \partial_{\mu} \rho_{\nu} - \partial_{\nu} \rho_{\mu}, \qquad
   F_{\mu\nu} = \partial_{\mu} A_{\nu} - \partial_{\nu} A_{\mu}, \quad
\end{equation}
and $m_{\omega}$, $m_{\rho}$ and $m_{\sigma}$ are the masses of 
$\omega$, $\rho$ and $\sigma$, respectively.  The $\sigma$ field has self-interactions
with strength parameters $b$ and $c$~\cite{Boguta:1977xi,Boguta:1981px,Pannert:1987jsc}. While the $\omega$, $\rho$, photon and $\sigma$
fields are to be introduced as mean field, the $\eta^{\prime}$ meson is introduced as 
a matter field with mass $m_{\eta^{\prime}}$ as well as the nucleon. The $\eta^{\prime}$ meson couples to the $\sigma$ field with the coupling strength $g_{\sigma\eta^{\prime}}$.
This $\eta^{\prime}$-$\sigma$ coupling induces the interaction between nucleon and the $\eta^{\prime}$ meson mediated by the $\sigma$ field as in the linear $\sigma$ model~\cite{Sakai:2013nba}. We assume no $\eta^{\prime}$-$\omega$ coupling according to the fact that the Weinberg-Tomozawa interaction vanishes for the $\eta^{\prime}N$ channel~\cite{Jido:2012av}. The Weinberg-Tomozawa interaction is explained by the vector meson exchange.
The value of the $\eta^{\prime}$-$\sigma$ coupling $g_{\sigma\eta^{\prime}}$ is the most important parameter in this work. It determines the strength of the attractive interaction between $\eta^{\prime}$ and nucleon and the depth of the $\eta^{\prime}$ potential in the nucleus.
Thus, the nature of $\eta^{\prime}$-nucleus bound systems, such as binding energies, is sensitive to the value of $g_{\sigma\eta^{\prime}}$. Here we fix it based on the result of the linear $\sigma$ model~\cite{Sakai:2013nba}.

We treat the $\sigma$, $\omega$, $\rho$ and $\gamma$ boson fields as static mean fields allowed
to have spatial variation. We assuming spherical nuclei, the mean fields have only 
radial dependence and the spacial components of the vector field vanish in the mean field 
approximation. The equations of motion for these fields read
\begin{subequations}
\label{eq:EqoM}
\begin{eqnarray}
   \bigl(-\nabla^2+m_\sigma^2\bigr)\sigma(r)&=&
   -bmg_\sigma^3\sigma(r)^2-cg_\sigma^4\sigma(r)^3 +  g_\sigma\rho_s(r)
    + g_{\sigma \eta^{\prime}} \frac{m_{\eta^{\prime}}}{E_{\eta^{\prime}}} \rho_{\eta^{\prime}}(r) ,
    \label{eq:sigmaEqoM} \\  
\bigl(-\nabla^2+m_\omega^2\bigr)\omega_0(r)&=&g_\omega(\rho_p(r)+\rho_n(r)),
      \label{eq:omegaEqoM} \\
\bigl(-\nabla^2+m_\rho^2\bigr)\rho_0(r)&=&\frac{1}{2}g_\rho(\rho_p(r)-\rho_n(r)),
       \label{eq:rhoEqoM}\\
-\nabla^2A_0(r)&=&e\rho_p(r),    \label{eq:photonEqoM}
\end{eqnarray}
\end{subequations}
with the nuclear scalar density $\rho_{s}(r)$, the proton number density $\rho_{p}(r)$,
the neutron number density $\rho_{n}(r)$, the $\eta^{\prime}$ number density $\rho_{\eta^{\prime}}(r)$
and the $\eta^{\prime}$ energy $E_{\eta^{\prime}}$.
Here we take the local density approximation, and these densities are given as functions of the radial variable.
These differential equations can be solved by introducing the Green function 
$G_{0}(\vec r, \vec r^{\,\prime})$ for 
\begin{equation}
(-\nabla^{2} + m^{2})G_{0}(\vec r, \vec r^{\,\prime}) =
\delta(\vec r - \vec r^{\,\prime}).
\end{equation} 
For a spherical source $\rho(r)$, the angular integral can be calculated, and 
the solution of the differential equation 
\begin{equation}
(-\nabla^{2} + m^{2})\phi(r) =\rho(r) ,
\end{equation}
with the boundary condition $\phi(r) \to 0$ for $r\to \infty$
is given by
\begin{equation}
    \phi(r) = \int_{0}^{\infty} G_{0}(r-r^{\prime}) \rho(r^{\prime}) dr^{\prime}, 
\end{equation}
with
\begin{equation}
    G_{0}(r-r^{\prime}) = \frac{1}{m} \frac{\sinh mr_{<}}{r_{<}} \frac{e^{-m r_{>}}}{r_{>}},
\end{equation}
where $r_{<} = r$, $r_{>} = r^{\prime}$ for $r< r^{\prime}$, 
and $r_{<} = r^{\prime}$, $r_{>} = r$ for $r^{\prime}< r$.
The mean fields can be obtained, once the nuclear and $\eta^{\prime}$ densities are given. 
For Eq.~\eqref{eq:sigmaEqoM}, we assume an initial distribution of the $\sigma$ mean field for the right hand side, and we solve Eq.~\eqref{eq:sigmaEqoM} with the Green function. We iterate this procedure until we obtain a self-consistent solution. 

We take the independent particle picture for the nucleon and the $\eta^{\prime}$ meson. 
These particles are moving independently in a one-body potential produced by the mean fields. 
The ground state of the nucleus is composed of the nucleons which fill lowest orbits according to  
Pauli principle. 
Because the mean fields are spherical, the one-body potentials are also spherical. Separating 
the radial and angular valuables, we introduce the wavefunction for the nucleon with 
total angular moment $j$ and magnetic quantum number $m$ as
\begin{equation}
   \psi_{jm}^{\pm}(r,\theta,\phi) 
   = \left( \begin{array}{c} f(r) {\cal Y}_{jm}^{\pm}(\theta,\phi) \\ i g(r) {\cal Y}_{jm}^{\mp}(\theta,\phi)
   \end{array} \right),
\end{equation}
where the superscript of the nucleon wavefunction corresponds to 
orbital angular momentum $\ell = j \mp \frac{1}{2}$ and ${\cal Y}_{jm}^{\pm}$ is the spinor spherical function
given by
\begin{eqnarray}
   {\cal Y}^{+}_{jm}(\theta, \phi) &=& \frac{1}{\sqrt {2j}} 
   \left(\begin{array}{c} 
   \sqrt{j+m}\, Y_{j-\frac{1}{2}}^{m-\frac{1}{2}}(\theta,\phi) \\
   \sqrt{j-m}\, Y_{j-\frac{1}{2}}^{m+\frac{1}{2}}(\theta,\phi) 
    \end{array} \right) , \\
   {\cal Y}^{-}_{jm}(\theta, \phi) &=& \frac{1}{\sqrt {2(j+1)}} 
   \left(\begin{array}{c} 
   -\sqrt{j-m+1}\, Y_{j+\frac{1}{2}}^{m-\frac{1}{2}}(\theta,\phi) \\
   \sqrt{j-m+1}\, Y_{j+\frac{1}{2}}^{m+\frac{1}{2}}(\theta,\phi) 
    \end{array} \right),
\end{eqnarray}
with the spherical harmonics $Y_{\ell}^{m}(\theta,\phi)$.  The Dirac equation 
for the nucleon is given for  $F(r) = r f(r)$ and $G(r) = rg(r)$, 
\begin{eqnarray}
  \left(\frac{d}{dr} -  \frac{\kappa}{r} \right) F(r) - \left[ E_{N} - V(r) + m - S(r) \right] G(r) &=& 0 ,
    \label{eq:FDirac}\\
  \left(\frac{d}{dr} + \frac{\kappa}{r}\right)G(r) + \left[ E_{N} - V(r) - m + S(r) \right] F(r) &=& 0 ,
      \label{eq:GDirac}
\end{eqnarray}
where $\kappa = \pm (j+\frac12)$, $E_{N}$ is the nucleon energy,
$V(r)$ is the vector potential and $S(r)$ is the scalar 
potential. These potentials are given by the mean fields as
\begin{eqnarray}
   V_{p} (r) &=& g_{\omega} \omega_{0}(r) + \frac{1}{2} g_{\rho} \rho_{0}(r) + e A_{0}(r) , \\
   S_{p} (r) &=&  g_{\sigma} \sigma(r),
\end{eqnarray}
for proton and
\begin{eqnarray}
   V_{n} (r) &=& g_{\omega} \omega_{0}(r) - \frac{1}{2} g_{\rho} \rho_{0}(r), \\
   S_{n} (r) &=&  g_{\sigma} \sigma(r),
\end{eqnarray}
for neutron. The nucleon wavefunctions and nucleon energy are obtained by solving the Dirac equation, 
once the density dependences of the mean fields are given. With the obtained wavefunctions,
we calculate the nuclear densities. The scalar density is calculated by
\begin{eqnarray}
   \rho_{s}(r) = \sum_{a}^{\rm occ.} \left(\frac{2j_{a}+1}{4\pi r^{2}}\right) \left[ |F_{a}(r)|^{2} - |G_{a}(r)|^{2}]\right],
\end{eqnarray}
where $j_{a}$ is the total angular moment of the state $a$ and
the summation is taken over the occupied states of the protons and neutrons, 
and the nuclear number density is obtained by
\begin{eqnarray}
   \rho_{N}(r) = \sum_{a}^{\rm occ.} \left(\frac{2j_{a}+1}{4\pi r^{2}}\right) \left[ |F_{a}(r)|^{2} + |G_{a}(r)|^{2}]\right],
\end{eqnarray}
where $N$ is proton or neutron and the summation is taken over the occupied states of proton or neutron. 
The nuclear 
densities are normalized as $\int \rho_{p}(r) d^{3}r = Z$ and $\int \rho_{n}(r) d^{3}r = N$
with the proton number $Z$ and the neutron number $N$.  
The effective mass of nucleon in the $\sigma$ mean field is defined as
\begin{equation} 
   m^{*}  = m - g_{\sigma} \sigma(r).
\end{equation}

In the similar way, we solve the Klein-Gordon equation for the $\eta^{\prime}$ meson. Introducing the
radial wavefunction as
\begin{equation}
    \eta^{\prime}(r,\theta,\phi) = \frac{R_{\eta^{\prime}}(r)}{r} Y_{\ell}^{m}(\theta,\phi),
\end{equation}
we have the radial Klein-Gordon equation
\begin{equation}
     \left( -  \frac{d^{2}}{dr^{2}}  + \frac{\ell(\ell+1)}{r^{2}} + m_{\eta^{\prime}}^{2} 
     -2g_{\sigma\eta^{\prime}} m_{\eta^{\prime}} \sigma(r) - E_{\eta^{\prime}}^{2}\right) R_{\eta^{\prime}}(r) 
     =0.   \label{eq:etaEqoM}
\end{equation}
Solving the above equation, we obtain the $\eta^{\prime}$ energy and wavefunction for 
an $\eta^{\prime}$ bound state with angular momentum $\ell$.
For the $\eta^{\prime}$ density, 
we take an average for the magnetic quantum number $m$
and obtain 
\begin{equation}
  \rho_{\eta^{\prime}}(r) = \frac{E_{\eta^{\prime}}}{4\pi r^{2}} |R(r)|^{2}.
\end{equation}
The wavefunction is normalized as $\int \rho_{\eta^{\prime}}(r) d^{3}r = 1$. 
The effective mass of the $\eta^{\prime}$ meson, $m_{\eta^{\prime}}^{*}$, is defined as
\begin{equation}
  m_{\eta^{\prime}}^{*2} = m_{\eta^{\prime}}^{2} - 2g_{\sigma\eta^{\prime}} m_{\eta^{\prime}} \sigma.
  \label{eq:etapEM}
\end{equation}
Thanks to the $\sigma$ mean field, the effective $\eta^{\prime}$ mass is reduced in the nuclear matter
for the attractive interaction with $g_{\sigma \eta^{\prime}}>0$.
For finite nuclei, 
this effect is seen as an attractive potential for the $\eta^{\prime}$ meson and 
one expects some bound states of the $\eta^{\prime}$ meson in the nucleus.  
Concerning the $\omega$-$\eta^{\prime}$ coupling, the possible source of the vector meson 
coupling might be the covariant derivative obtained by replacing derivative to $\partial_{\mu} + i g \omega_{\mu}$.
But, because the $\eta^{\prime}$ meson is a neutral particle, there is no such a gauge-type vector meson coupling. 
A possible coupling may be so-called anomalous coupling like 
$\epsilon_{\mu\nu\rho\sigma} \partial^{\mu}\omega^{\nu}\partial^{\rho}\omega^{\sigma} \eta^{\prime}$ \cite{Bijnens:1988kx,Bramon:1997va}. However, owing to the antisymmetric nature of the $\omega$ fields, one of the $\omega$ fields should be its spatial component and it should vanish in a spherical nucleus. Thus, it is natural that the $\omega$-$\eta^{\prime}$ coupling as a source of repulsive interaction be absent in spherical nuclei.
Some studies \cite{Collins:2017vev,Anisovich:2018yoo,Tiator:2018heh} suggest that there are nucleon resonances which couple to $\eta^{\prime}N$. If the $\eta^{\prime}$ meson has strong couplings to these resonances in the nucleus, this can be a source of the nuclear absorption of the $\eta^{\prime}$ meson into the nucleus.
Such resonance effects can be implemented as an external complex potential of equation of motion for $\eta^{\prime}$ given in Eq.~\eqref{eq:etaEqoM}. Alternatively if one regards such resonances as effective constituents of the system,
one needs a coupled channel calculation between the $\eta^{\prime}$-nucleus and $N^{*}$-nucleus systems.

To obtain a self-consistent solution of Eqs.~\eqref{eq:EqoM}, 
\eqref{eq:FDirac}, \eqref{eq:GDirac}, \eqref{eq:etaEqoM}, 
we first provide an initial condition for the nuclear number densities
and the scalar density as, for instance, a Woods-Saxon type distribution. 
With this initial condition for the densities, 
we solve the mean field equations and obtain the distributions of the mean fields.
For the $\sigma$ mean field, we assume an initial distribution of the $\sigma$ field
and solve Eq.~(\ref{eq:sigmaEqoM}) self-consistently. 
With the mean field distributions, we have the one-body potentials for the nucleon and the $\eta^{\prime}$
meson. Then, we solve the Dirac and Klein-Gordon equations for the nucleon and the $\eta^{\prime}$ meson
and calculate their density distributions.  With these densities, we solve the mean field equations again.
We iterate these steps until we obtain a self-consistent solution. 
To construct $\eta^{\prime}$ mesonic nuclei, 
we first solve the equations without the $\eta^{\prime}$ meson and obtain a self-consistent solution 
for a usual nucleus. We take this configuration as the initial condition for the $\eta^{\prime}$ mesonic 
nuclei, because the $\eta^{\prime}$ mesonic nucleus is produced from 
an existing nucleus in formation experiments. 

We also calculate the total energy of the system
\begin{equation}
{E_{\rm tot} = E_{\rm mean} + E_{\rm nuc} + E_{\eta^{\prime}}}, \label{eq:Etot}
\end{equation}
where the mean field energy $E_{\rm mean}$ and the nucleon energy $E_{\rm nuc}$ are defined by
\begin{eqnarray}
   E_{\rm mean}&=& 4\pi \int_{0}^{\infty} dr\, r^{2} \left\{ 
   \frac12 \left[ \left(\nabla \sigma(r)\right)^{2} + m_{\sigma}^{2} \sigma(r)^{2} \right]
   - \frac12 \left[ \left(\nabla \omega_{0}(r)\right)^{2} + m_{\omega}^{2} \omega_{0}(r)^{2} \right]
   \right.\nonumber \\ && \left.
   - \frac12 \left[ \left(\nabla \rho_{0}(r)\right)^{2} + m_{\rho}^{2} \rho_{0}(r)^{2} \right]
   - \frac12 \left(\nabla A_{0}(r)\right)^{2} 
   + \frac13 b m g_{\sigma}^{3} \sigma(r)^{3} + \frac14 c g_{\sigma} \sigma(r)^{4}
   \right\} \label{eq:Emean}
   \\ 
   E_{\rm nuc}&=&
   \sum_{\alpha}^{\rm occ.} (2j_{a}+1) E_{p,\alpha}
   + \sum_{\alpha}^{\rm occ.} (2j_{a}+1) E_{n,\alpha}, \label{eq:Enuc}
\end{eqnarray}
respectively.
We also confirm whether the obtained solution is stable against the energy. 

\begin{table}[t]
\begin{center}
\caption{\label{ta:mass} Hadron masses in units of MeV. Isospin symmetry is assumed.}
\begin{tabular}{ccccc}\hline
$m$ & $m_{\eta^{\prime}}$ & $m_{\sigma}$ & $m_{\omega}$ &  $m_{\rho}$ \\
\hline
938.92 & 957.78 & 450& 775.5 & 770 \\
\hline
\end{tabular}
\end{center}
\end{table}

\begin{table}[t]
\begin{center}
\caption{\label{ta:parameters}
Parameters of the model. These parameter sets reproduce
the saturation density $\rho_{0}=0.153$ fm$^{-3}$, the binding energy per nucleon $B/A=-16.3$ MeV
and the symmetry energy $A_{\rm sym}=32.5$ MeV.  Each parameter set provides a different 
compressibility $K$ and nucleon effective mass $m^{*}$. 
The masses of the mean fields are $m_{\sigma}=450$ MeV,
$m_{\omega}=775.5$ MeV and $m_{\rho} = 770$ MeV.  
These parameters were originally determined in Ref.~\cite{CompactStars}. 
}
\begin{tabular}{c|ccccccccc}
\hline
No.&1&2&3&4&5&6&7&8&9\\
\hline
${g_\sigma}$& 8.12 &7.67 &7.18 &7.98 &7.47 &6.89 &7.83 &7.28 &6.61 \\
${g_\omega}$& 10.43&9.31 &8.03&	10.43&9.31&8.03 &10.43 &9.31 &8.03 \\
${g_\rho}$& 8.25 &8.48 &8.68 &8.25 &8.48 &8.68 &8.25 &8.48 &8.68 \\
$b$ $(\times 100)$ & 0.561& 0.878 &1.460 &0.431 &0.628 &0.880 &0.295 &0.360 &0.248 \\
$c$ $(\times 100)$ &-0.699 &-1.010&-1.241&-0.410 &-0.341 &0.692 &-0.107 &0.372 	&2.800\\
\hline
$K$ [MeV] & 199.1 &200.0 &200.0 &249.8 &249.9 &249.9 &299.7 &299.8 &300.0 \\
${m^*}/{m}$ &0.700&0.750&0.800&0.700&0.750&0.800&0.700&0.750&0.800\\
\hline
\end{tabular}
\end{center}
\end{table}

The parameters of the mean fields are determined so as to reproduce the nuclear matter properties. 
The formulation for the calculation of the nuclear matter properties is summarized in Appendix~\ref{sec:NM}. 
The masses of the hadrons in this study are shown in Table~\ref{ta:mass}. We assume isospin symmetry. 
The mass of the $\sigma$ field determines the surface properties of finite nuclei. 
We use the parameter sets shown in textbook~\cite{CompactStars}, where several parameter sets 
were proposed. There parameter sets reproduce 
the saturation density $\rho_{0}=0.153$ fm$^{-3}$, the binding energy per nucleon $B/A=-16.3$ MeV
and the symmetry energy $A_{\rm sym}=32.5$ MeV, while the values of the compressibility $K$ and
the effective mass of nucleon at the saturation density, $m^{*}$, depend on the parameter sets. 
It is known that the current mean field approach, in which the $\sigma$ field has cubic and quadratic 
self-interactions, provides negative coefficients for the quadratic interaction, when smaller compressibility 
and effective nucleon mass are to be reproduced~\cite{Maruyama:1993jb}.
In Table~\ref{ta:parameters} we show the parameter sets used in this study and the reproduced 
compressibility $K$ and the nucleon effective mass $m^{*}$. 
In Fig.~\ref{fig:saturation}, we show the equation of state for the symmetric nuclear matter
reproduced with these parameter sets. This figure implies that all of the parameter sets well reproduce
the saturation property and agree each other for lower densities $\rho < 0.2$ fm$^{-3}$, 
while in higher densities $\rho > 0.2$ fm$^{-3}$ different equations of state are predicted. 
In this work, we compare $\eta^{\prime}$ mesonic nuclei produced by these different nuclear matters.

\section{Results}

In this section, we show our numerical results on the $\eta^{\prime}$ mesonic nuclei. 
Our target nuclei in this work are double magic nuclei, $^{16}$O and $^{40}$Ca, and $^{12}$C.
These are spherical stable nuclei and may be described well in the relativistic mean field theory. 
The $^{16}$O nucleus is composed of eight protons and neutrons, and in the ground state these 
nucleons occupy the $1s_{1/2}$, $1p_{3/2}$ and $1p_{1/2}$ orbits. 
The $^{40}$Ca nucleus has twenty protons and neutrons and they are in the 
$1s_{1/2}$, $1p_{3/2}$, $1p_{1/2}$, $1d_{5/2}$, $1d_{3/2}$ and $2s_{1/2}$ states.  
The $^{12}$C nucleus has six protons and neutrons in the $1s_{1/2}$ and $1p_{3/2}$ orbits. 
First of all, we show that the present approach relevantly describes normal nuclei, and then
we discuss the results of the $\eta^{\prime}$ mesonic nuclei. 

\subsection{Normal nuclei without $\eta^{\prime}$}

\begin{figure}[tb]
  \subfloat[$^{16}$O]{
  \includegraphics[width=0.48\textwidth]{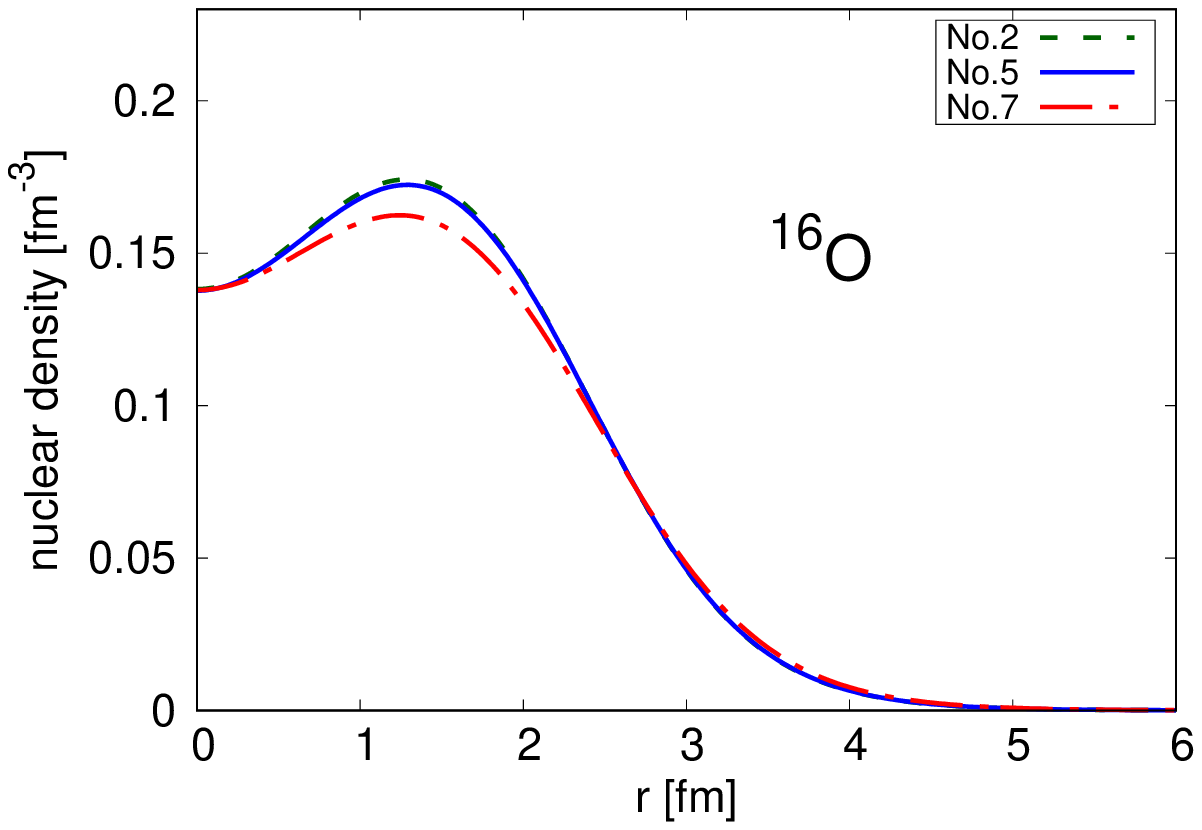}
  }
  \subfloat[$^{40}$Ca]{
  \includegraphics[width=0.48\textwidth]{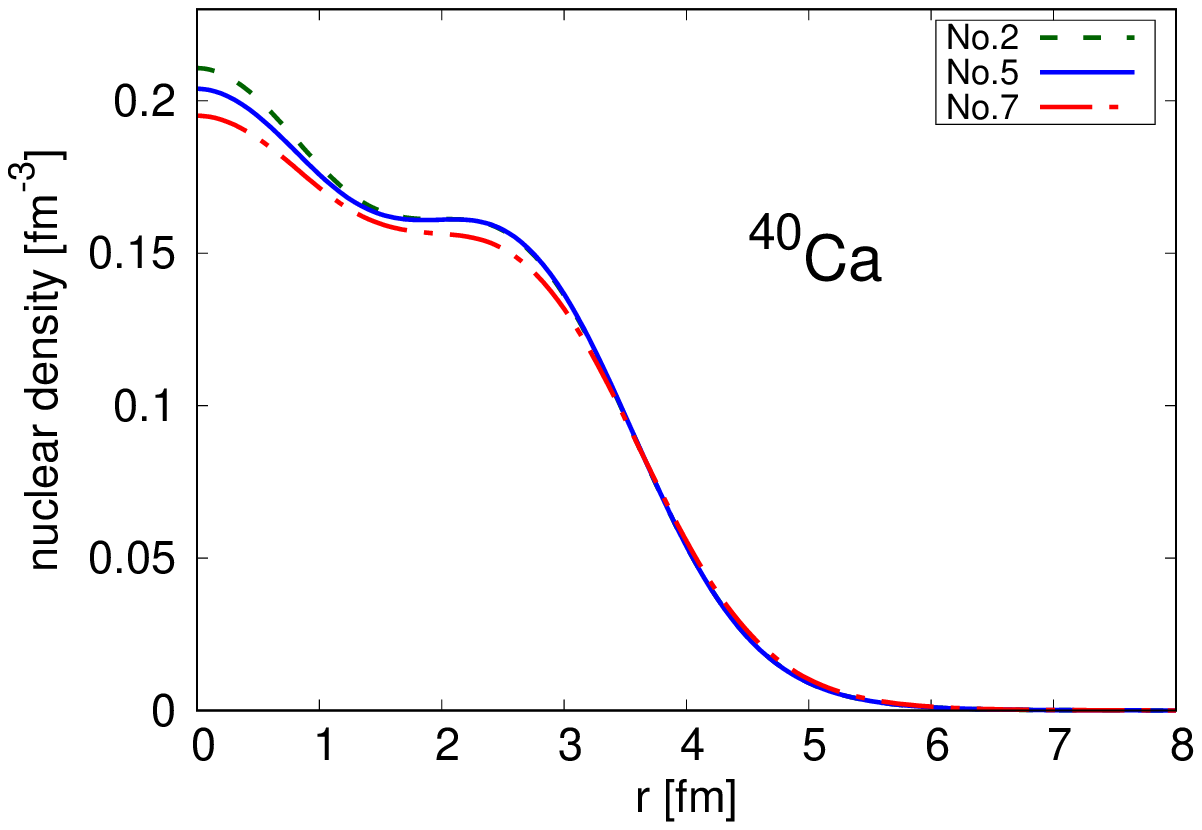}
  }\\
  \subfloat[$^{12}$C]{
  \includegraphics[width=0.48\textwidth]{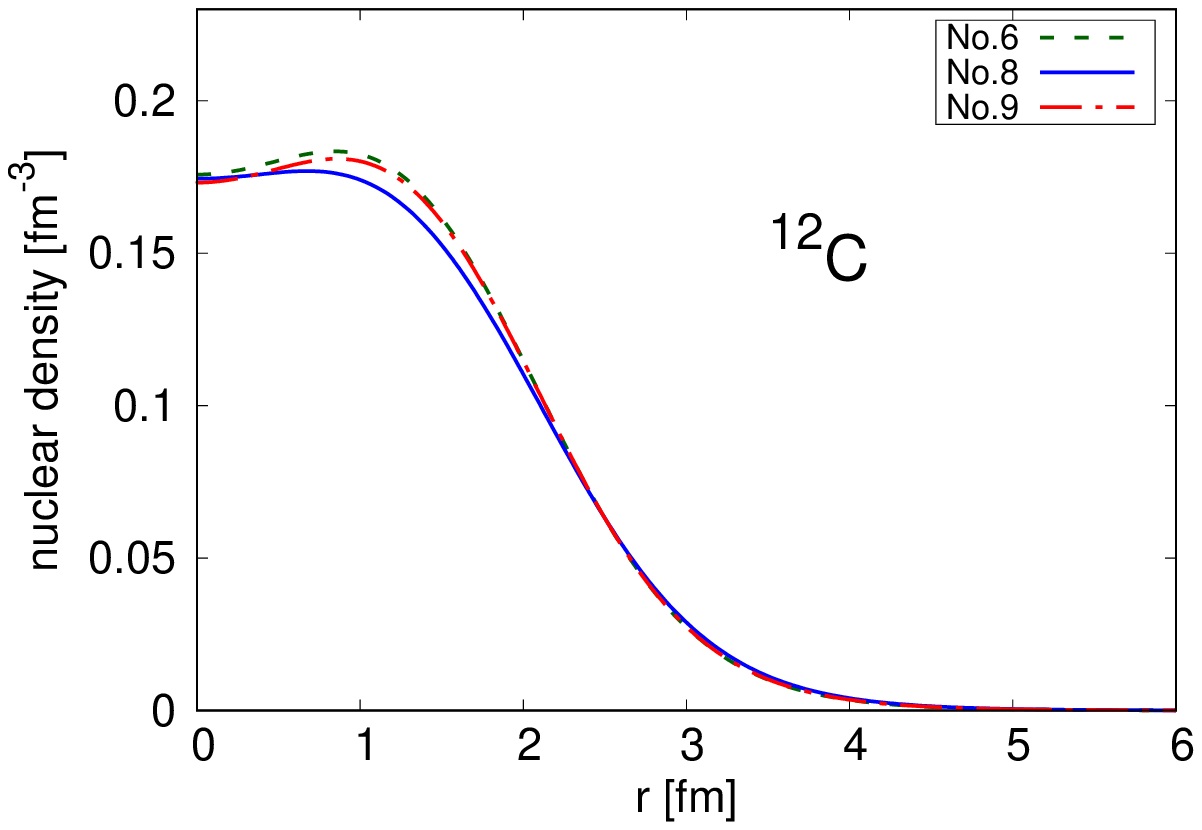}
  }
\caption{Nuclear densities obtained in present model. \label{fig:rhoN}}
\end{figure}

To construct finite nuclei, we introduce initial nuclear density and $\sigma$ field configuration,
which are assumed to be
Woods-Saxon type distributions with an appropriate radius for the 
nuclear density and the $\sigma$ configuration. Then, we solve the equations of motion
for the mean fields and the nucleons till a self-consistent solution is obtained. For these nuclei,
the stabilized solution can be obtained within several iterations. 
In Fig.~\ref{fig:rhoN}, we plot the nuclear density calculated in the present approach 
for (a) $^{16}$O, (b) $^{40}$Ca and (c)~$^{12}$C. 
Here we show the results of the parameter sets of 2, 5, 7 for $^{16}$O and 
$^{40}$Ca and the parameter sets of 6, 8, 9
for $^{12}$C. These parameter sets reproduce the
equivalent saturation property but provide different equations of state especially at higher densities. 
Parameter set 2 and 7 supply soft and hard nuclear matter in which the ground state energies 
at higher density are lower and higher, respectively. Parameter set 5 produces a medium 
equation of state. We do not find any marked differences among the choice of the parameter. 

In Table~\ref{tab:normal_nucleus}, we show the numerical results of the nuclear properties 
obtained in the present model. For the binding energy per particle, $E_{\rm tot}/A - m_{N}$, 
we obtain around $-8$ MeV for these nuclei, which is a typical binding energy for stable nuclei. 
We also show the average of the nucleon binding energy, $E_{\rm nuc}/A - m_{N}$, which 
is around $-20$ MeV, and the mean field energy per particle, $E_{\rm mean}/A$, which is repulsive 
with $14$ MeV for these nuclei. Owing that we have the saturation properties of the nuclear matter 
in this model, these energies are slightly dependent on the nucleus. 
For the $^{12}$C nucleus, we take model 6, 8 and 9. These models have a positive value of the parameter $c$ and are used in the calculation of
the $\eta^{\prime}$ mesonic nucleus. 

\begin{table}
\begin{center}
\caption{  \label{tab:normal_nucleus}
Numerical results of the nuclear properties obtained in the present model in units of MeV. The energies of the mean field and the nucleons, 
$E_{\rm mean}$ and $E_{\rm nuc}$,  are
defined in Eqs.~\eqref{eq:Emean} and \eqref{eq:Enuc}, respectively, and
the total energy of the nucleus $E_{\rm tot}$ is given by $E_{\rm tot} = E_{\rm mean} + E_{\rm nuc}$. 
Here we show the binding energy per particle, $E_{\rm tot}/A - m_{N}$, 
the average nucleon binding energy, $E_{\rm nuc}/A - m_{N}$ and the mean field energy per particle,
$E_{\rm mean}/A$, where $A$ is the mass number of the nucleus. 
}
\begin{tabular}{l|D{.}{.}{0}D{.}{.}{0}D{.}{.}{1}|D{.}{.}{0}D{.}{.}{0}D{.}{.}{1}|D{.}{.}{0}D{.}{.}{0}D{.}{.}{1}}
\hline
 \multicolumn{1}{c|}{nucleus} & \multicolumn{3}{c|}{$^{16}$O}
 & \multicolumn{3}{c|}{$^{40}$Ca}& \multicolumn{3}{c}{$^{12}$C}\\
 \multicolumn{1}{c|}{model} &  \multicolumn{1}{c}{2} &   \multicolumn{1}{c}{5} &   \multicolumn{1}{c|}{7}
 &  \multicolumn{1}{c}{2} &   \multicolumn{1}{c}{5} &   \multicolumn{1}{c|}{7}
 &  \multicolumn{1}{c}{6} &   \multicolumn{1}{c}{8} &   \multicolumn{1}{c}{9} \\
 \hline
 $E_{\rm tot}/A - m_{N}$ & -7.7 & -7.1 & -5.9  & -8.6 & -8.1 & -7.2 & -5.9 & -4.7 & -5.0 \\
 $E_{\rm nuc}/A - m_{N}$ & -21.8 & -21.2 & -20.1 & -22.7 & -22.3 & -21.9 & -20.0 & -19.0 & -19.2\\
 $E_{\rm mean}/A$ & 14.0 & 14.1 & 14.2 & 14.1 & 14.2 & 14.7& 14.1 & 14.2 & 14.2 \\
 \hline
\end{tabular}
\end{center}
\end{table}

\subsection{$\eta^{\prime}$ mesonic nuclei}

First of all, we determine the strength of the $\eta^{\prime}$-$\sigma$ coupling. 
We make a good use of the result obtained in the linear $\sigma$ model~\cite{Sakai:2013nba}.
In Ref.~\cite{Sakai:2013nba}, under the assumption that partial restoration of chiral symmetry takes 
place in the nuclear matter with 30\% reduction of the magnitude of the quark condensate at the 
saturation density, one finds about 80 MeV mass reduction of the $\eta^{\prime}$ meson at 
the saturation density
\footnote{The original work~\cite{Sakai:2013nba} uses 700 MeV for the $\sigma$ mass, while we take 450 MeV for the $\sigma$ mass to reproduce the surface properties of nuclei. It is known that the value of the $\sigma$ mass is not so important in the linear $\sigma$ model and the model is fixed by the other parameters, such as the decay constant of pion and the masses of $\pi$, $K$, $\eta$ and $\eta^{\prime}$. We have checked that even with 450 MeV for the $\sigma$ mass one obtains a similar linear $\sigma$ model, in which about 80 MeV mass reduction of the $\eta^{\prime}$ mass is obtained at the saturation density.}. 
We determine the $g_{\sigma\eta^{\prime}}$ coupling constant 
such that the effective $\eta^{\prime}$ mass $m_{\eta^{\prime}}^{*}$
given in Eq.~\eqref{eq:etapEM} is 80 MeV smaller than the in-vacuum mass at the saturation 
density in our model, that is, $\rho_{0} = 0.153$ fm$^{-3}$. To calculate the effective $\eta^{\prime}$ 
mass, we consider the infinite symmetric nuclear matter with the $\sigma$ and $\omega$ mean fields
and calculate the value of the $\sigma$ mean field at the saturation density. With this value
we determine the coupling constant $g_{\sigma \eta^{\prime}}$ from Eq.~\eqref{eq:etapEM}. 
The determined parameters are listed in Table~\ref{ta:eta}. 
We see in the table that the $\sigma\eta^{\prime}$ coupling constants are about one third of the $\sigma N$ coupling constant.

\begin{table}[tb]
\begin{center}
\caption{\label{ta:eta} Coupling constants of $\sigma$-$\eta^\prime$ interaction. 
These are determined so as to reproduce the effective $\eta^{\prime}$ mass 80 MeV smaller than
the in-vacuum mass at the saturation density.  }
\begin{tabular}{c|ccccccccc}
\hline
No.&1&2&3&4&5&6&7&8&9\\
\hline
$g_{\sigma\eta^{\prime}}$&2.21&2.51&2.94&2.17&2.44&2.82&2.13&2.38&2.70\\
\hline
\end{tabular}
\end{center}
\end{table}

We solve the Klein-Gordon equation~\eqref{eq:etaEqoM} with a definite angular momentum 
for $\eta^{\prime}$ and obtain the $\eta^{\prime}$ energy $E_{\eta^{\prime}}$ for each 
angular momentum.  We also calculate radial excitation states if exist.  
We do not consider nuclear excited states and nucleons are in the ground state of 
the $\eta^{\prime}$-nucleus bound system.

\begin{figure}[tb]
  \centering
  \includegraphics[width=0.6\textwidth]{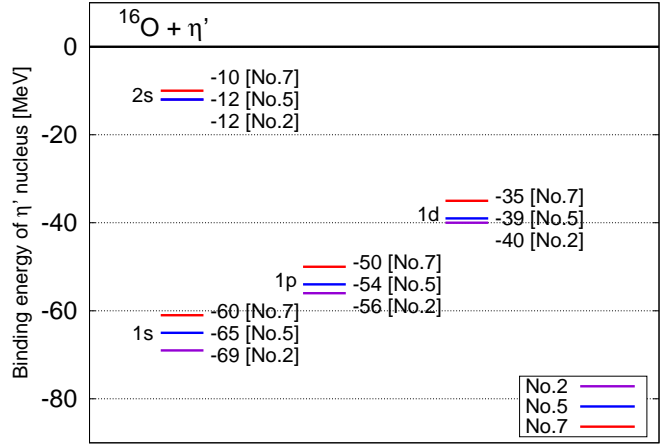}
  \caption{
  Binding energy spectrum of 
  the $\eta^{\prime}$-$^{16}$O system measured from the $\eta^{\prime} + ^{16}$O threshold.
  The binding energy is defined as $E_{B} = E_{\rm tot}- E_{^{16}{\rm O}}-m_{\eta^{\prime}}$
  with the value of the normal $^{16}$O energy, $E_{^{16}{\rm O}}$, 
  taken from our calculation for the normal nucleus.
  The letter appearing in the left side of each energy level shows the quantum number
  $n_{r}\ell$ of the $\eta^{\prime}$ state with the radial quantum number $n_{r}$ and the 
  orbital angular momentum $\ell$, while the number appearing in the right
  represents the binding energy in units of MeV.
  The binding energies 
  are calculated with three parameter sets. Parameter set 2 and 7 provide soft and hard nuclear media
  in higher densities, respectively, and parameter set 5 obtains a medium nuclear matter.  }
  \label{gr:Ospec}
\end{figure}

The energy spectrum of the $\eta^{\prime}$-$^{16}$O system is shown in Fig.~\ref{gr:Ospec}.
Here we show the binding energy defined by $E_{B} = E_{\rm tot}- E_{^{16}{\rm O}}-m_{\eta^{\prime}}$,
in which $E_{\rm tot}$ is the total energy of the system calculated as shown in Eq.~\eqref{eq:Etot}
and $E_{^{16}{\rm O}}$ is the total energy of the normal $^{16}$O calculated in the current model
without the $\eta^{\prime}$ meson. 
We find four bound states for $\eta^{\prime}$, the $1s$, $1p$, $1d$ and $2s$ states.  
In Table~\ref{tab:eta16O}, we show the details of the energy contents of the $\eta^{\prime}$ mesonic
nucleus. There we show the binding energy $E_{B}$ measured from the threshold of $^{16}{\rm O}+\eta^{\prime}$,
the $\eta^{\prime}$ binding energy, $E_{\eta^{\prime}} - m_{\eta^{\prime}}$,
the average of the nucleon binding energy, $E_{\rm nuc}/A - m_{N}$, and 
the mean field energy per particle, $E_{\rm mean}/(A+1)$, with $A=16$ for $^{16}$O.

It is interesting noting that for the case of the $\eta^{\prime}$ meson bound in the $s$ states
the total binding energy is smaller than the $\eta^{\prime}$ binding energy in magnitude, while
for the higher $\ell$ states the total binding energy is larger than the $\eta^{\prime}$ binding 
energy in magnitude.
This implies that, first of all, the nuclear modification effects are substantially large and should not be 
ignored. In particular, for the $1d$ state the $\eta^{\prime}$ binding energy is about 10 MeV, but
the total binding energy is as large as about 40 MeV. The difference stems from larger
binding energies of the nucleons in magnitude compared with the normal nucleus.  
With the $\eta^{\prime}$ meson in the nucleus, thanks to the strong $\eta^{\prime}$-$\sigma$ coupling,
the $\sigma$ mean field gets together to the $\eta^{\prime}$ meson and attracts the nucleons. 
Thus, the binding energies of the nucleons get larger in magnitude.  
The attraction also makes the nuclear density higher. As seen later, the central nuclear density 
for the mesonic nucleus with the $1s$ $\eta^{\prime}$ state gets about 1.5 larger than the 
normal density. This higher density makes the mean field energy enhanced repulsively. This 
is the reason that the bound states with the $\eta^{\prime}$ meson in the $s$ states has
a smaller binding energy than the $\eta^{\prime}$ binding energy in magnitude.

We find the model dependence that parameter set 2 provides
a deeper bound state, while parameter set 7 predicts a relatively shallower bound state than the others.
The model dependence stems from the behavior of the nuclear matter 
in higher densities. Parameter set 2 provides a softer nuclear matter.
There higher density nuclear matter has less energy per nucleon as shown in Fig.~\ref{fig:saturation}, and 
thus the matter obtained by parameter set 2 gets dense more easily than others. 
In higher density the $\sigma$ mean field is larger and the $\eta^{\prime}$ mass is suppressed better.
This provides larger attractive potential for $\eta^{\prime}$ in the nucleus.

\begin{table}
\begin{center}
\caption{  \label{tab:eta16O}
Numerical results of the properties of the $\eta^{\prime}$-$^{16}$O bound system obtained 
in the present model for each $\eta^{\prime}$ bound state in units of MeV. 
The binding energy $E_{B}$ measured from the threshold of $^{16}{\rm O}+\eta^{\prime}$
is defined as $E_{B} = E_{\rm tot}- E_{^{16}{\rm O}}-m_{\eta^{\prime}}$, where the value of 
the normal $^{16}$O energy, $E_{^{16}{\rm O}}$, is taken from our calculation for the normal nucleus. 
We show here also the $\eta^{\prime}$ binding energy, $E_{\eta^{\prime}} - m_{\eta^{\prime}}$,
the average of the nucleon binding energy, $E_{\rm nuc}/A - m_{N}$, and 
the mean field energy per particle, $E_{\rm mean}/(A+1)$, where $A=16$ for $^{16}$O.
Nuclear excited stats are not considered here and nucleus stays in the ground state.  }
\begin{tabular}{l|D{.}{.}{1}D{.}{.}{1}D{.}{.}{1}|D{.}{.}{1}D{.}{.}{1}D{.}{.}{1}}
\hline
 \multicolumn{1}{c|}{$\eta^{\prime}$ state} & \multicolumn{3}{c|}{$1s$}
 & \multicolumn{3}{c}{$1p$}\\
 \multicolumn{1}{c|}{model} &  \multicolumn{1}{c}{2} &   \multicolumn{1}{c}{5} &   \multicolumn{1}{c|}{7}
 &  \multicolumn{1}{c}{2} &   \multicolumn{1}{c}{5} &   \multicolumn{1}{c}{7} \\
 \hline
 $E_{B}$ & -69.2&-65.0&-60.5 &-56.2&-54.2&-50.0  \\
 $E_{\eta^{\prime}} - m_{\eta^{\prime}}$ & -84.0&-73.5	&-67.8 &-41.3&-39.2&-35.5 \\
 $E_{\rm nuc}/A - m_{N}$ &-26.1&-24.5&	-23.5 &-25.4&-24.4&-23.3\\
 $E_{\rm mean}/(A+1)$ & 18.2&16.8&17.0 &15.7&15.4&15.5 \\
 \hline
\hline
 \multicolumn{1}{c|}{$\eta^{\prime}$ state} & \multicolumn{3}{c|}{$1d$}
 & \multicolumn{3}{c}{$2s$}\\
 \multicolumn{1}{c|}{model} &  \multicolumn{1}{c}{2} &   \multicolumn{1}{c}{5} &   \multicolumn{1}{c|}{7}
 &  \multicolumn{1}{c}{2} &   \multicolumn{1}{c}{5} &   \multicolumn{1}{c}{7} \\
 \hline
 $E_{B}$ &-40.1&-39.1&-35.4 &-12.1&-11.6&-10.0\\
 $E_{\eta^{\prime}} - m_{\eta^{\prime}}$ &-12.7&-12.2&-9.9&-13.2&-12.4&-10.6 \\
 $E_{\rm nuc}/A - m_{N}$ &-24.5&-23.8&-22.7&-24.0&-23.3&-22.2\\
 $E_{\rm mean}/(A+1)$ &14.2&14.1&14.2&15.4&15.3&15.3 \\
 \hline
\end{tabular}
\end{center}
\end{table}

\begin{figure}[tb]
  \centering
  \subfloat[1s]{
  \includegraphics[width=0.45\textwidth]{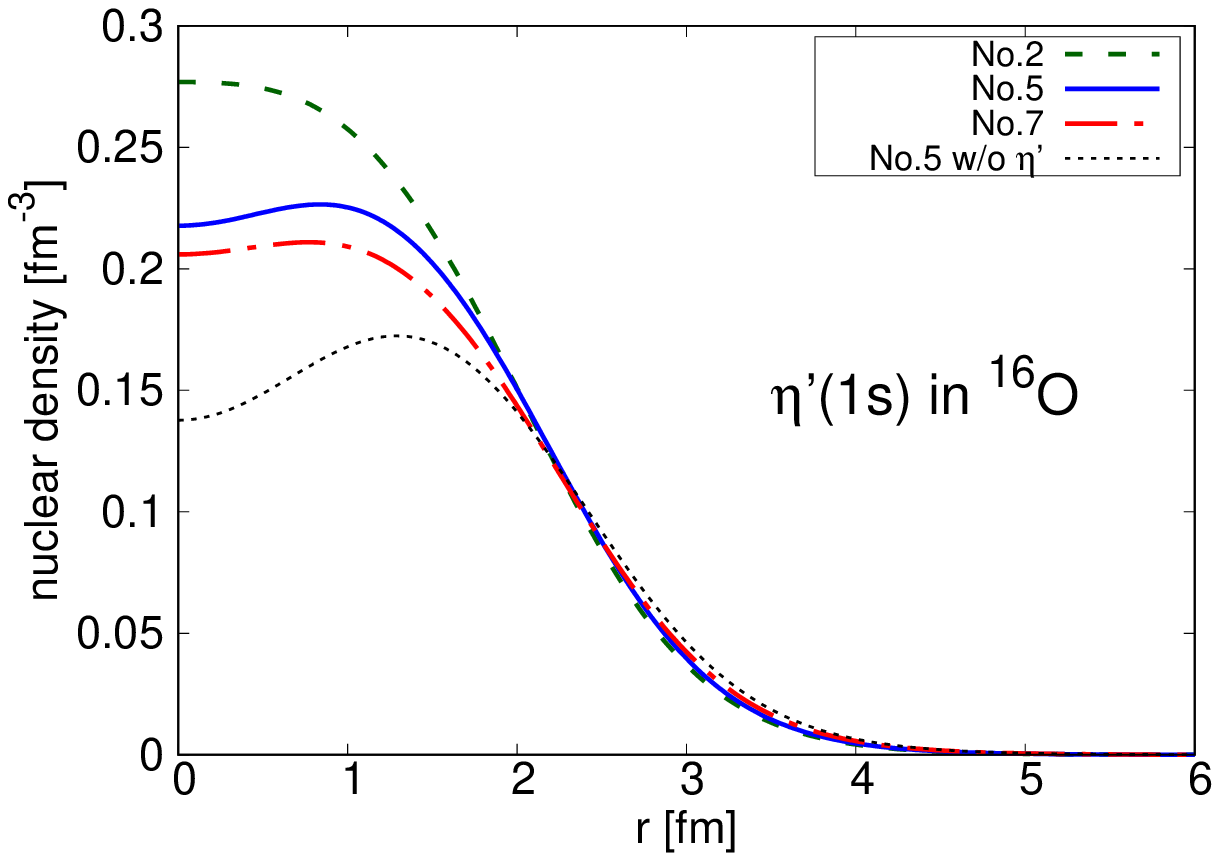}
  }
  \subfloat[1p]{
  \includegraphics[width=0.45\textwidth]{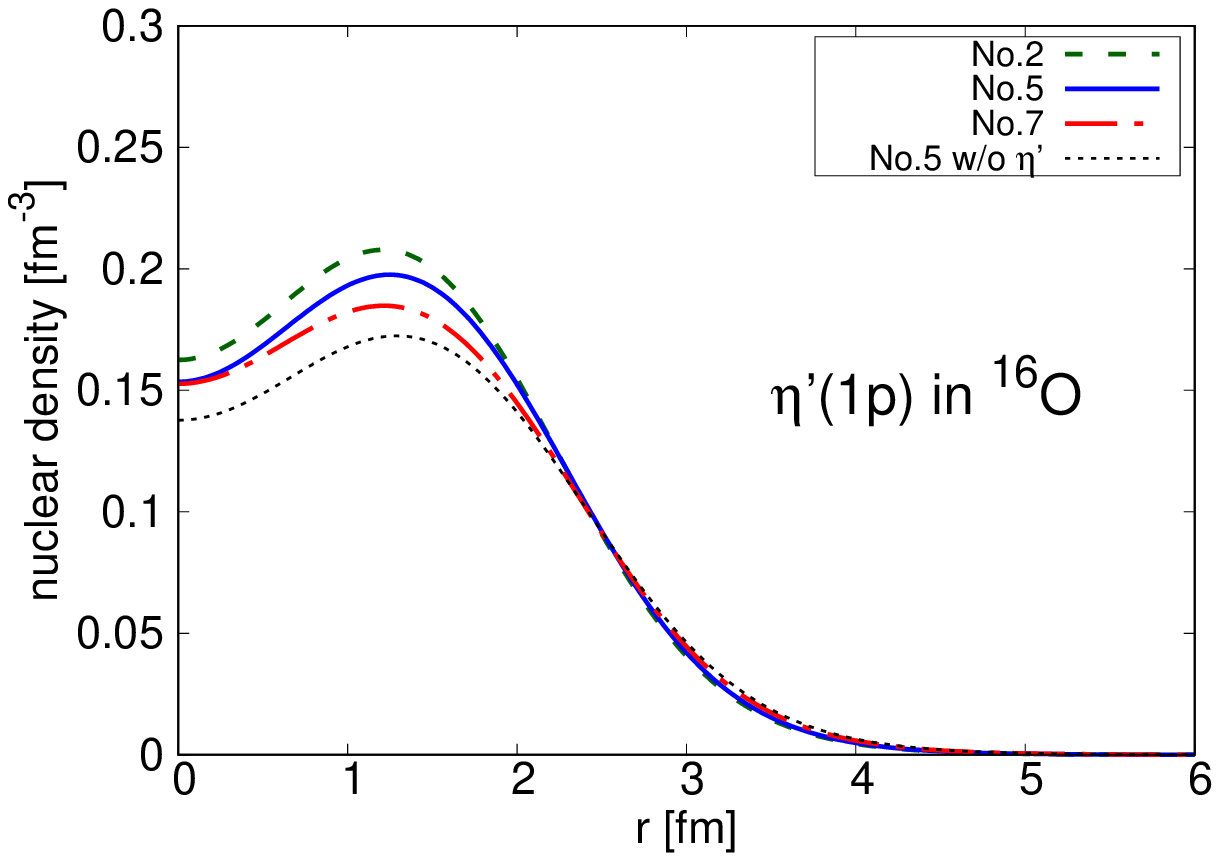}
  }\\
  \subfloat[2s]{
  \includegraphics[width=0.45\textwidth]{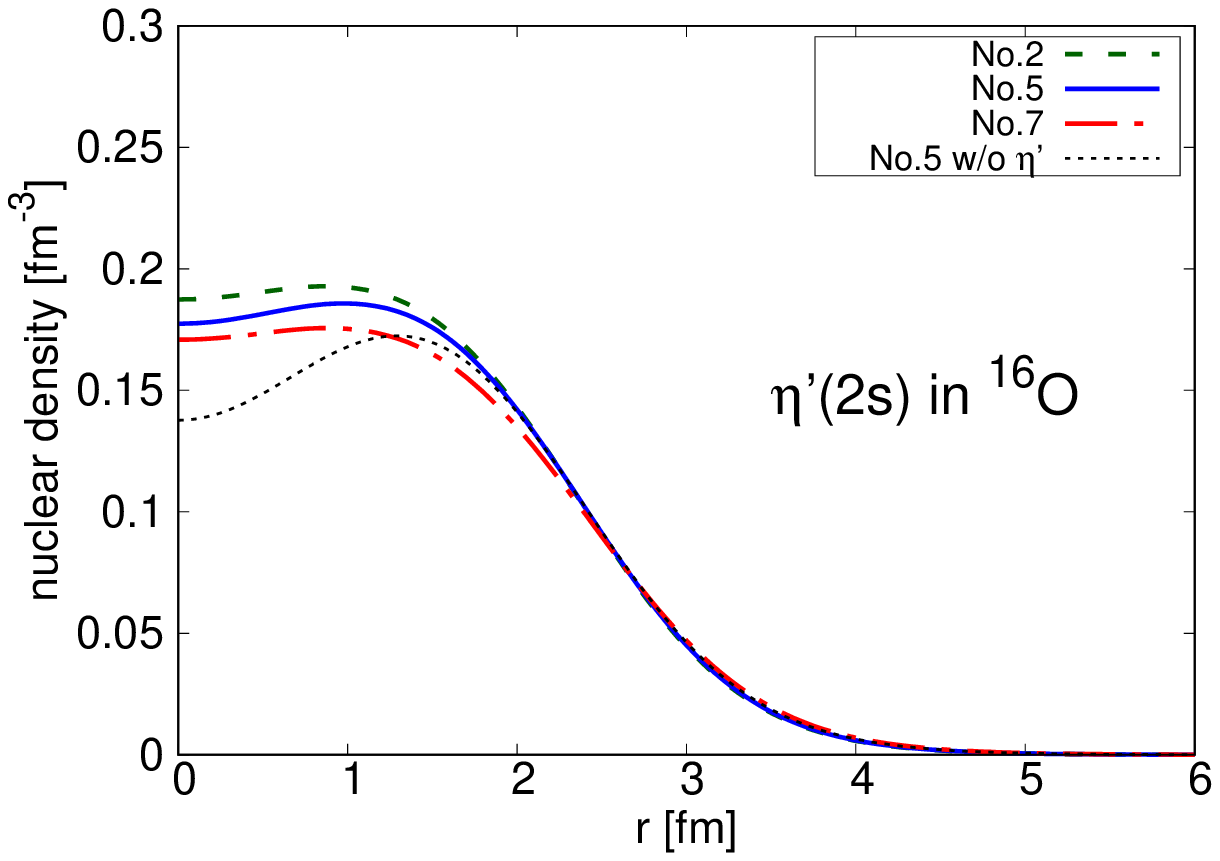}
  }
  \subfloat[1d]{
  \includegraphics[width=0.45\textwidth]{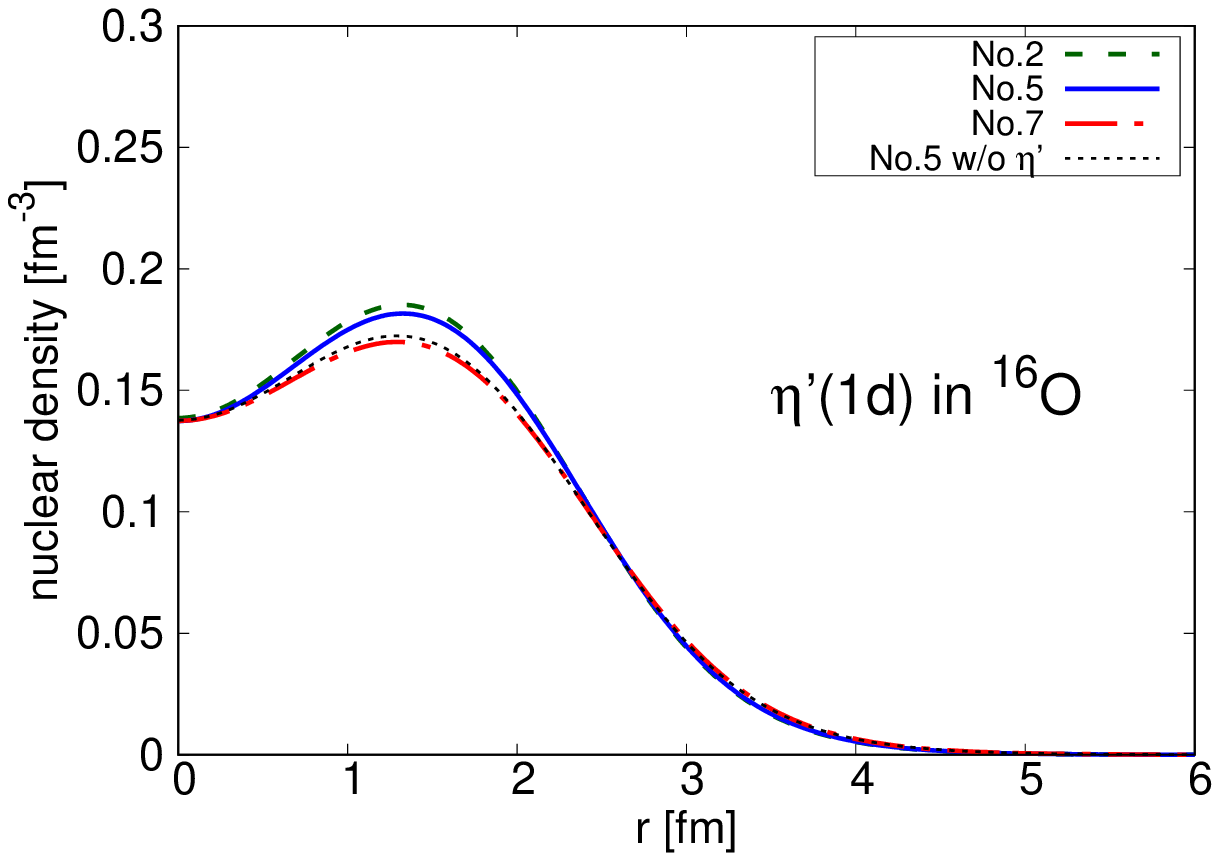}
  }
  \caption{Nuclear density profiles of the $\eta^{\prime}$ bound systems in $^{16}$O. The dotted line
  stands for the density distribution of the normal nucleus $^{16}$O without $\eta^{\prime}$ calculated
  with parameter set~5.  \label{gr:Odens}}
\end{figure}

\begin{figure}[tb]
  \centering
  \includegraphics[width=0.6\textwidth]{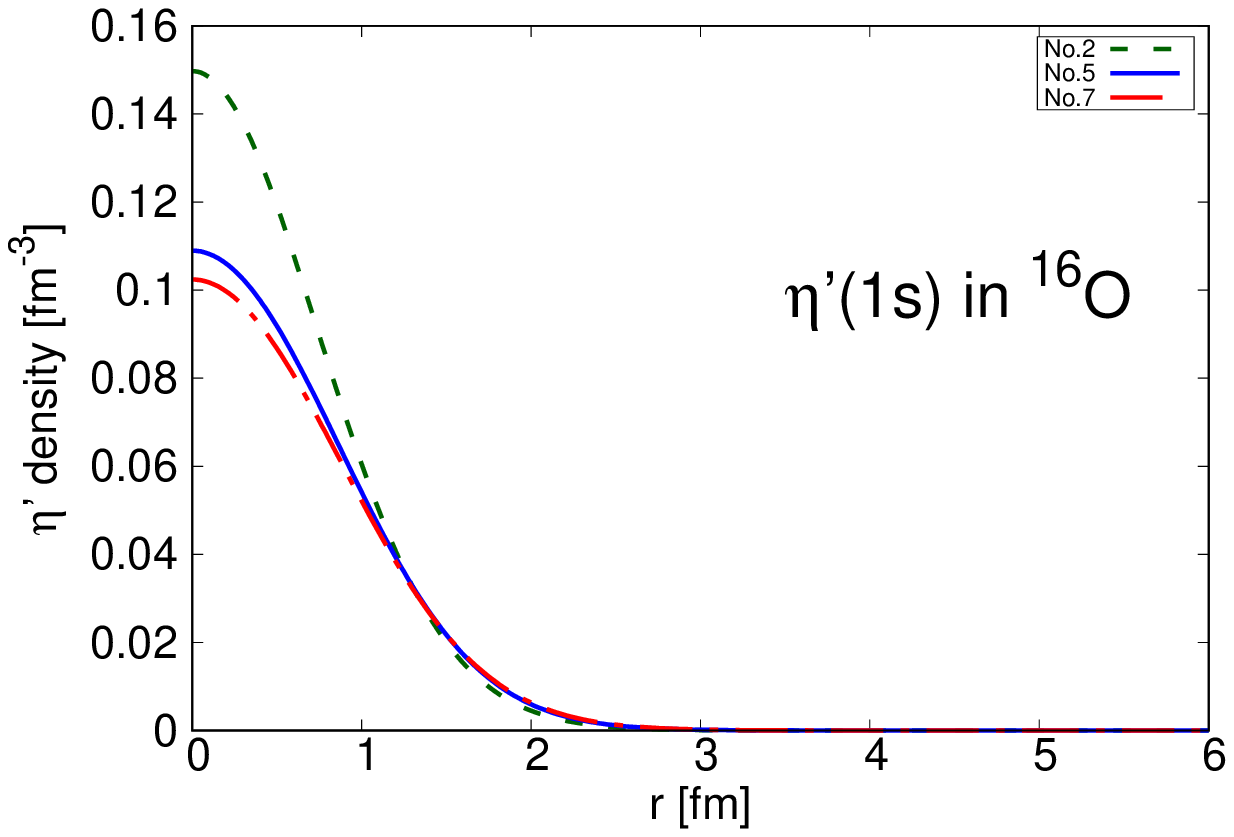}
  \caption{Density of $\eta^{\prime}$ in the $1s$ orbit of the $\eta^{\prime}$ mesonic nucleus 
  for $^{16}$O. }
  \label{gr:etap1sO}
\end{figure}

Figure \ref{gr:Odens} shows the nuclear density profiles for the $\eta^{\prime}$ bound systems in 
$^{16}$O. In this figure we also plot the density distribution of the normal nucleus $^{16}$O
calculated with parameter set 5 in the dashed lines for comparison.
For the case of the $\eta^{\prime}$ 1$s$ state, the central density gets 
1.5 to 2.0 times larger than the normal density. This is because, 
due to the $s$-wave nature the $\eta^{\prime}$ wave function is concentrated in the center of the nucleus, 
the attractive interaction between $\eta^{\prime}$ and $\sigma$ makes 
the $\sigma$ mean field gathered at the center, and, as a consequence, the nuclear density 
gets higher at the center of the nucleus. 
It is also notable that, because the the density distribution for the $1s$ bound state 
is quite different from the usual nuclear density distribution, 
the overlap integral for the wave functions of these states may be strongly suppressed 
and the production cross section of the $1s$ $\eta^{\prime}$ mesonic nucleus
in the nuclear reaction may be small. This implies that such deeper states
have disadvantage to be observed in experiments. 
For the $\eta^{\prime}$ excited states, the nuclear density distributions are very similar 
to the normal density distribution. 
In Fig.~\ref{gr:etap1sO}, we show also the density distribution of the $\eta^{\prime}$ meson 
for the $1s$ bound state. This shows that the $\eta^{\prime}$ density at the center of 
the nucleus is as larger as 0.1 - 0.15 fm$^{-3}$. In appendix~\ref{etapNM}, we discuss 
the nuclear matter properties in finite $\eta^{\prime}$ density.

\begin{figure}[tb]
  \centering
  \includegraphics[width=0.7\textwidth]{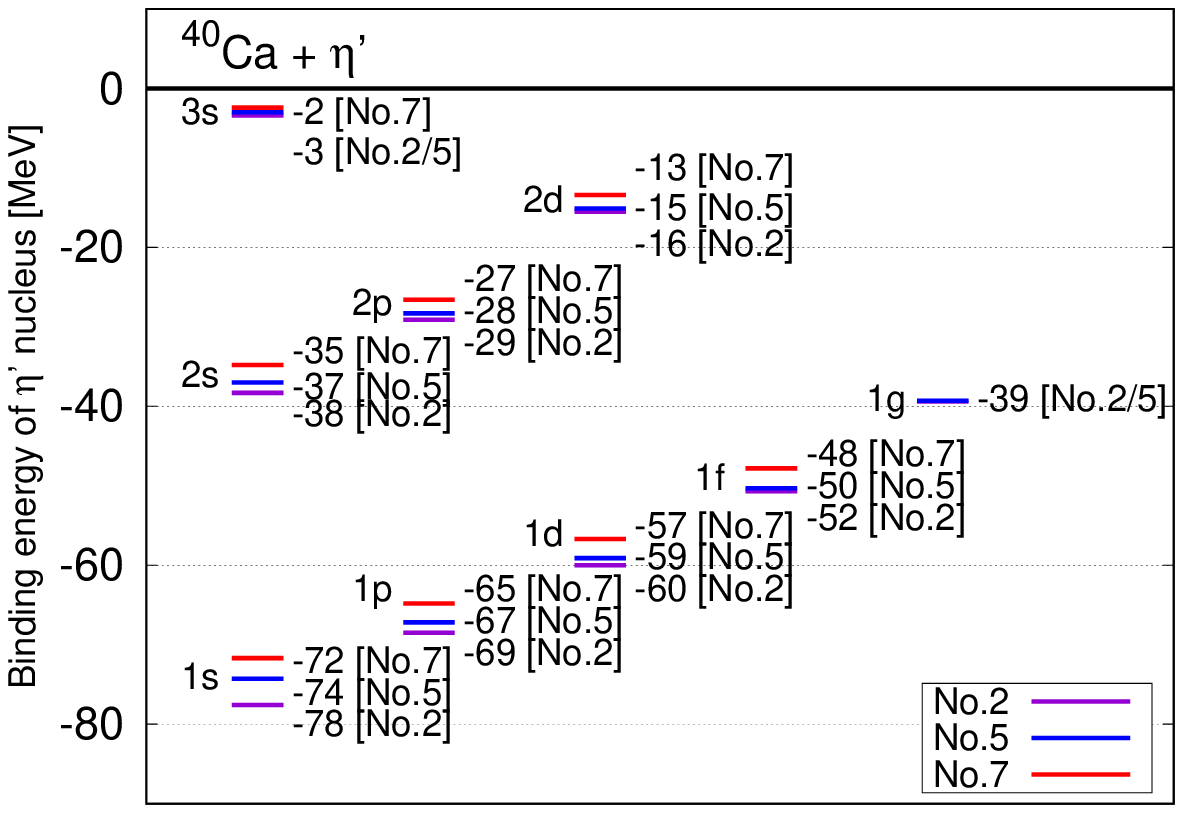}
  \caption{Binding energy spectrum of 
  the $\eta^{\prime}$-$^{40}$Ca system measured from the $\eta^{\prime} + ^{40}$Ca threshold.
  The binding energy is defined as $E_{B} = E_{\rm tot}- E_{^{40}{\rm Ca}}-m_{\eta^{\prime}}$
  with the value of the normal $^{40}$Ca energy, $E_{^{40}{\rm Ca}}$, 
  taken from our calculation for the normal nucleus.
  The letter appearing in the left side of each energy level shows the quantum number
  $n_{r}\ell$ of the $\eta^{\prime}$ state with the radial quantum number $n_{r}$ and the 
  orbital angular momentum $\ell$, while the number appearing in the right
  represents the binding energy in units of MeV.
  The binding energies 
  are calculated with three parameter sets. Parameter set 2 and 7 provide soft and hard nuclear media
  in higher densities, respectively, and parameter set 5 obtains a medium nuclear matter.  }
  \label{gr:Caspec}
\end{figure}

\begin{figure}[t]
  \centering

  \subfloat[1s]{
  \includegraphics[width=0.32\textwidth]{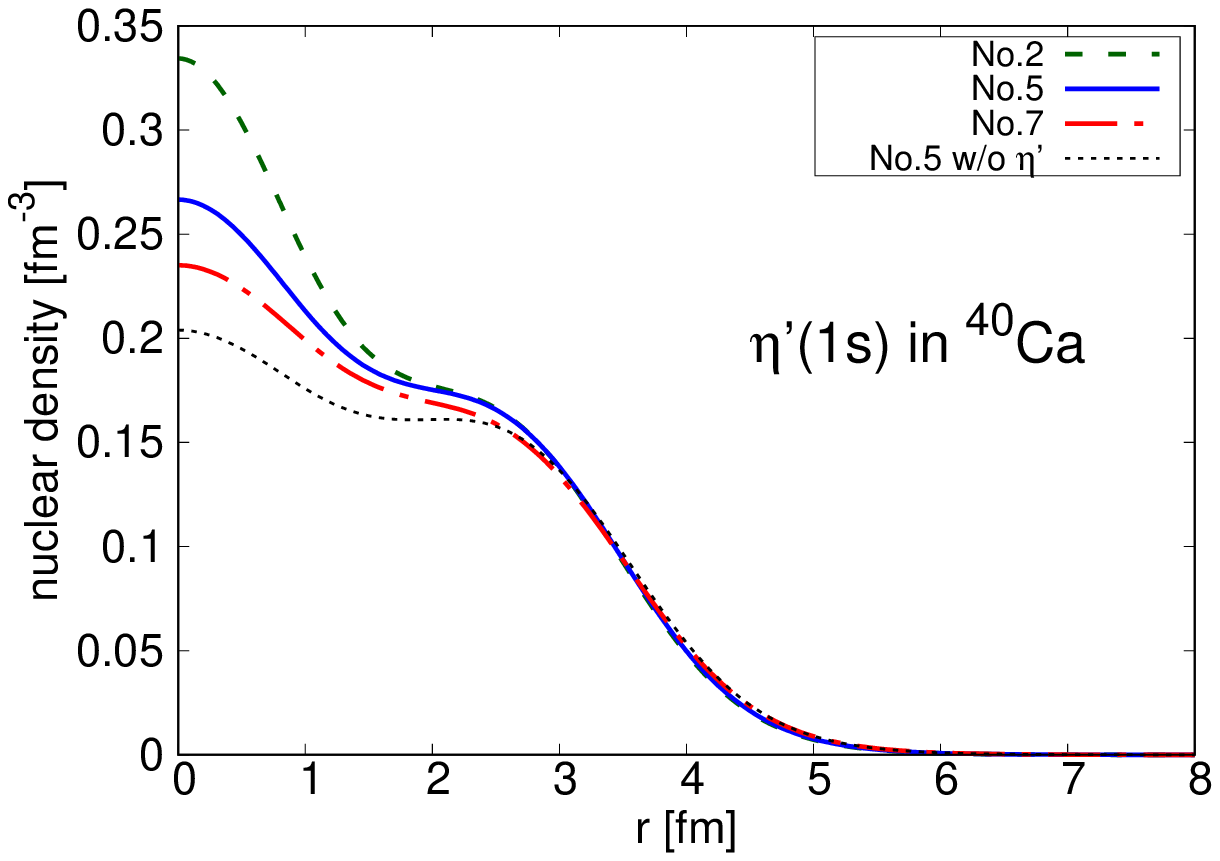}
  }
  \subfloat[1p]{
  \includegraphics[width=0.32\textwidth]{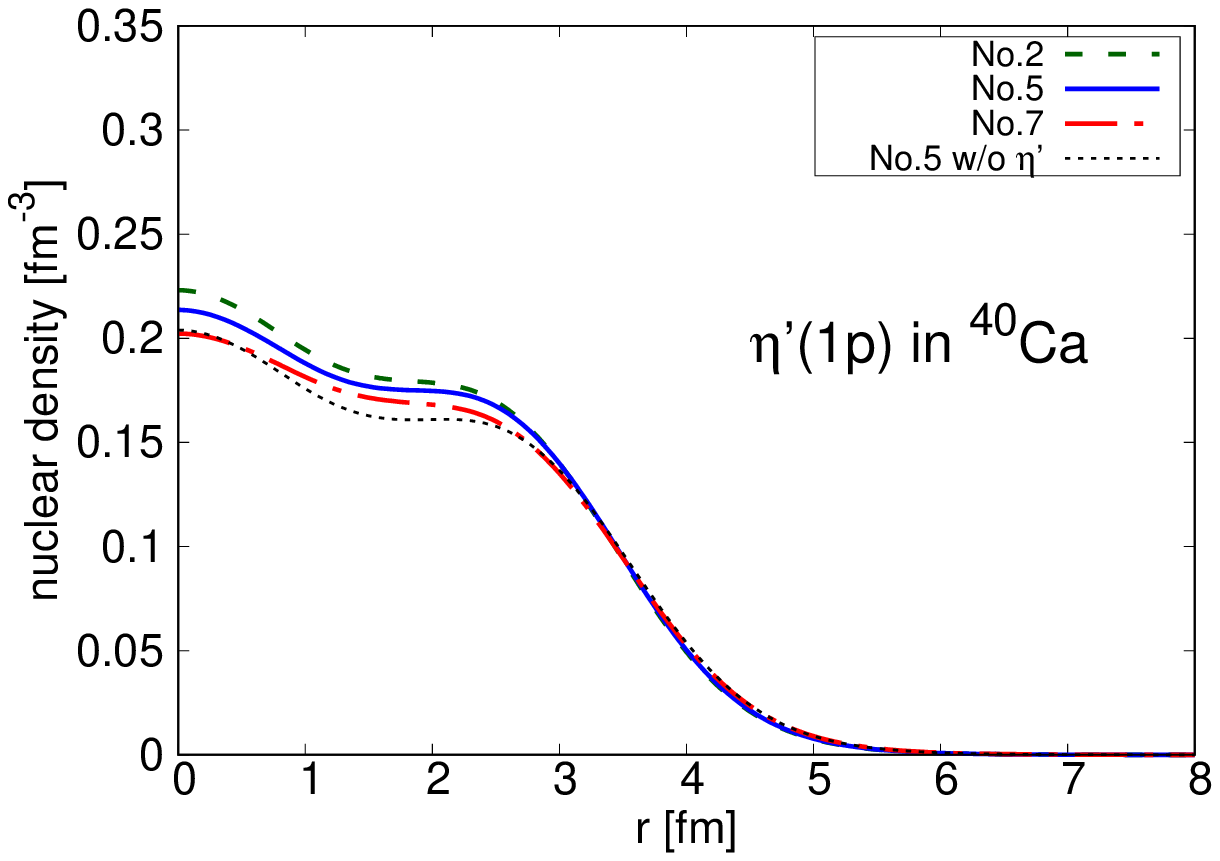}
  }
  \subfloat[1d]{
  \includegraphics[width=0.32\textwidth]{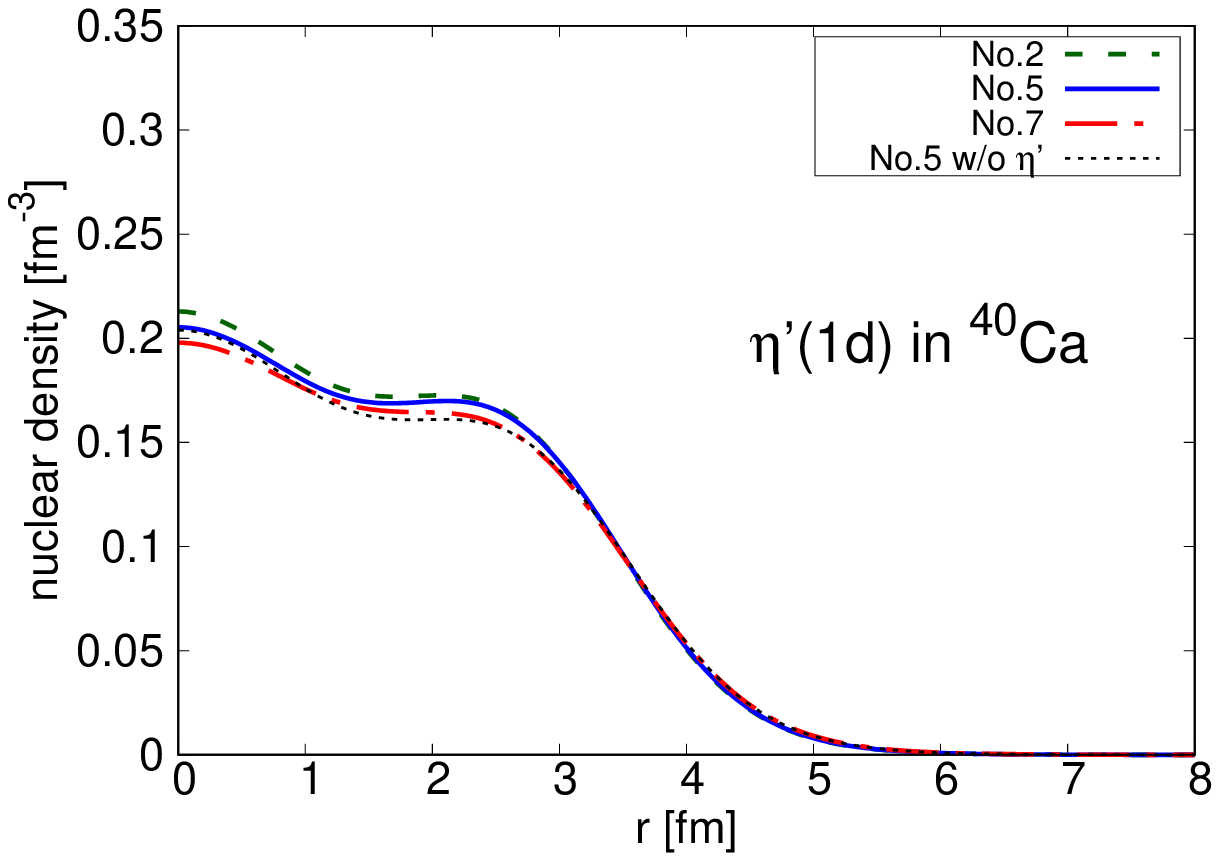}
  }\\

  \subfloat[2s]{
  \includegraphics[width=0.32\textwidth]{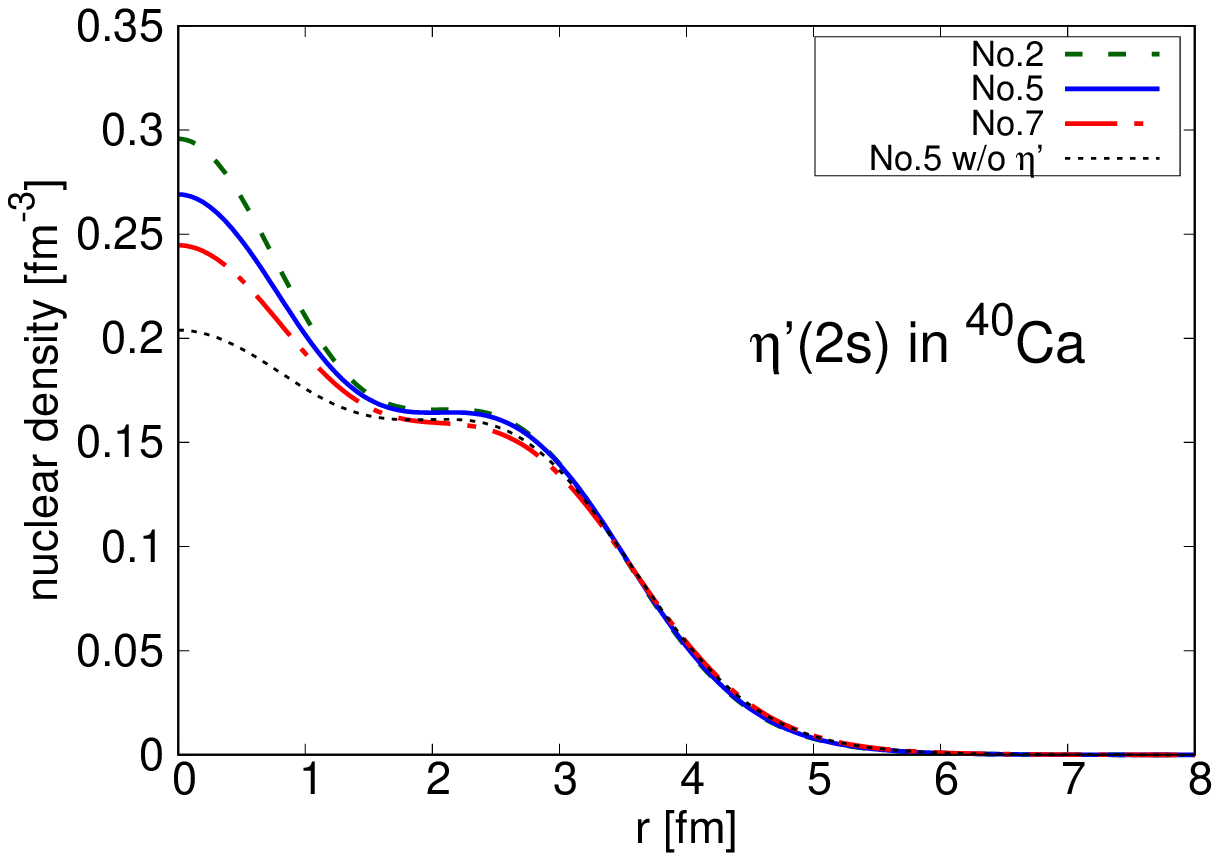}
  }
  \subfloat[2p]{
  \includegraphics[width=0.32\textwidth]{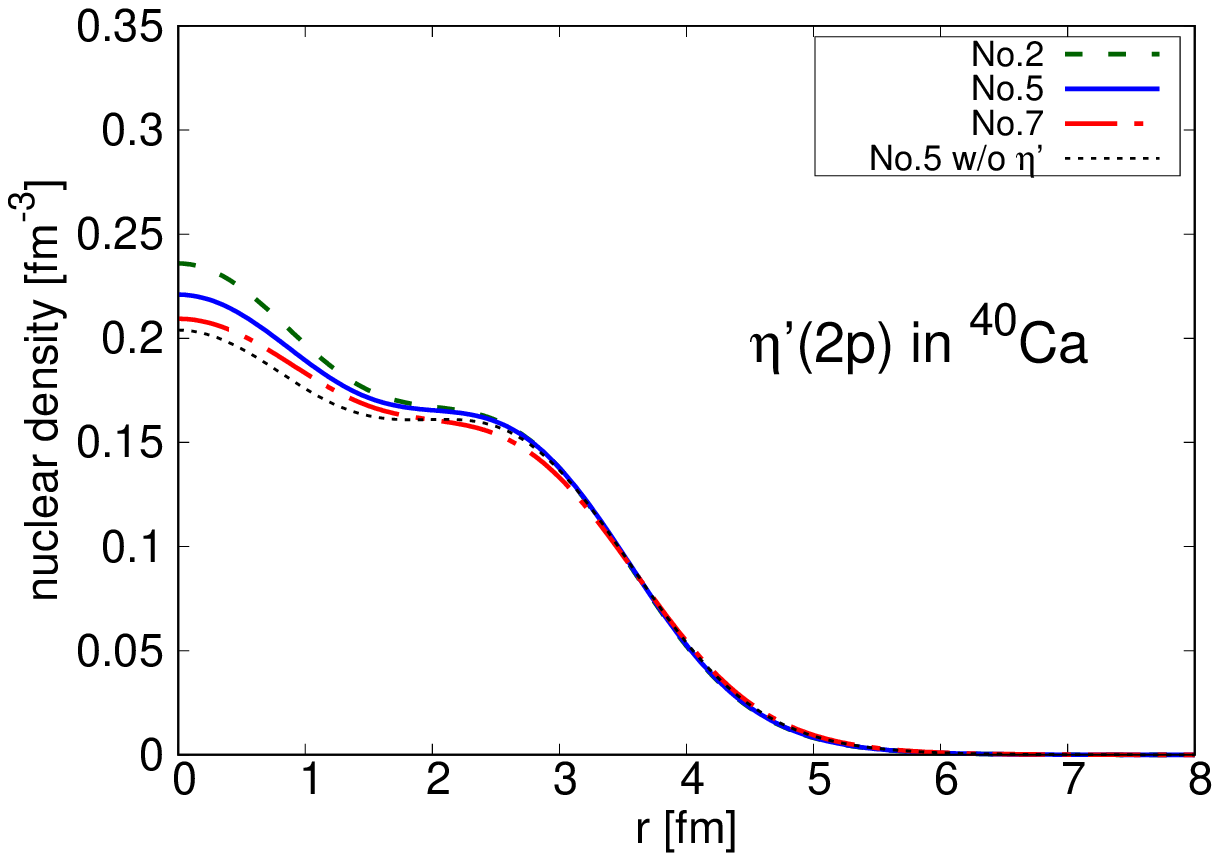}
  }
  \subfloat[2d]{
  \includegraphics[width=0.32\textwidth]{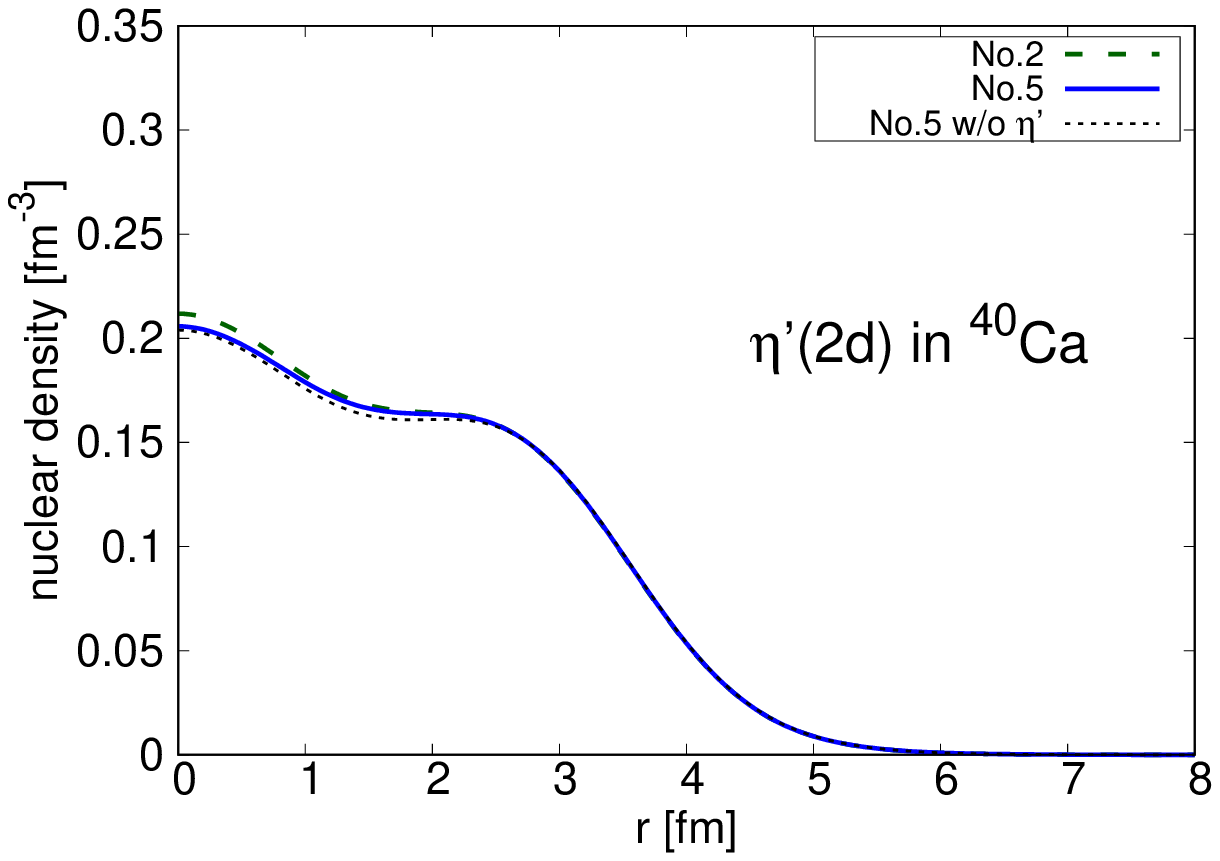}
  }\\

  \subfloat[3s]{
  \includegraphics[width=0.32\textwidth]{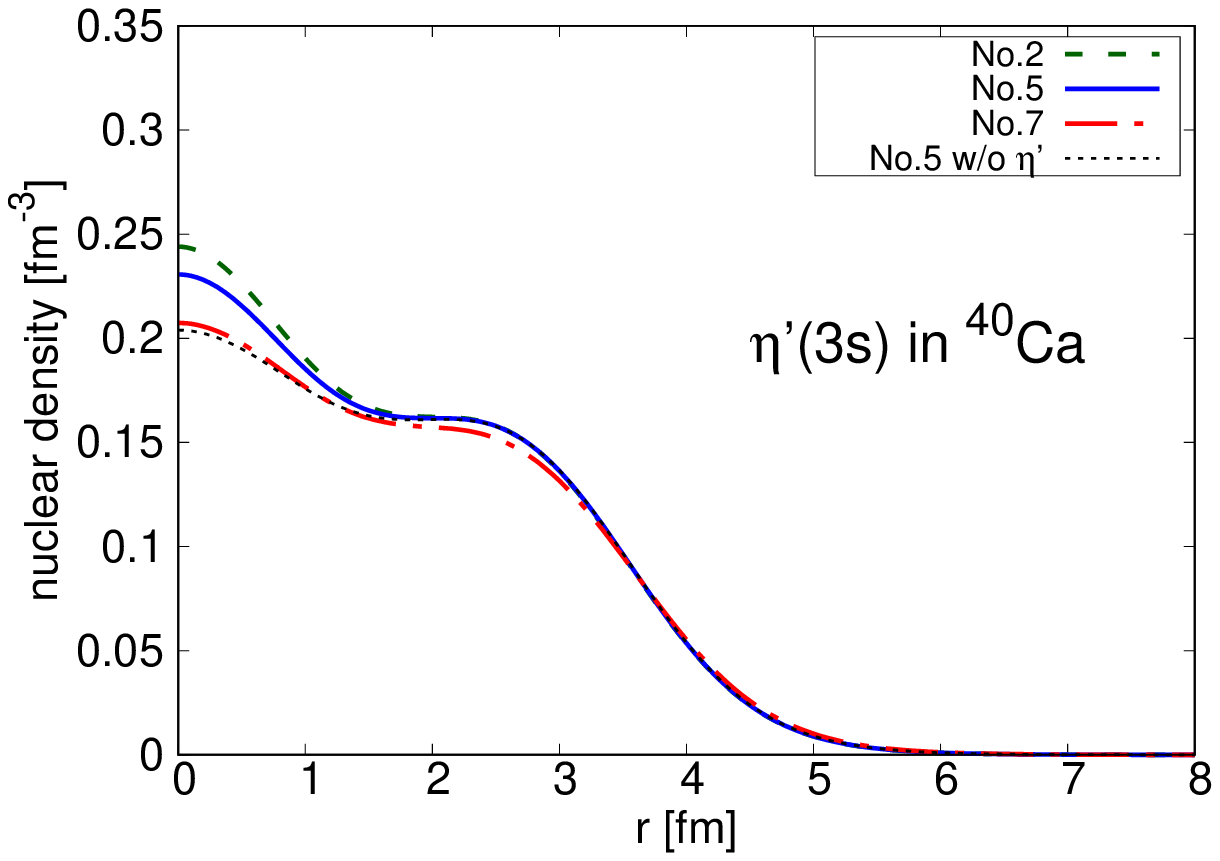}
  }
  \subfloat[1f]{
  \includegraphics[width=0.32\textwidth]{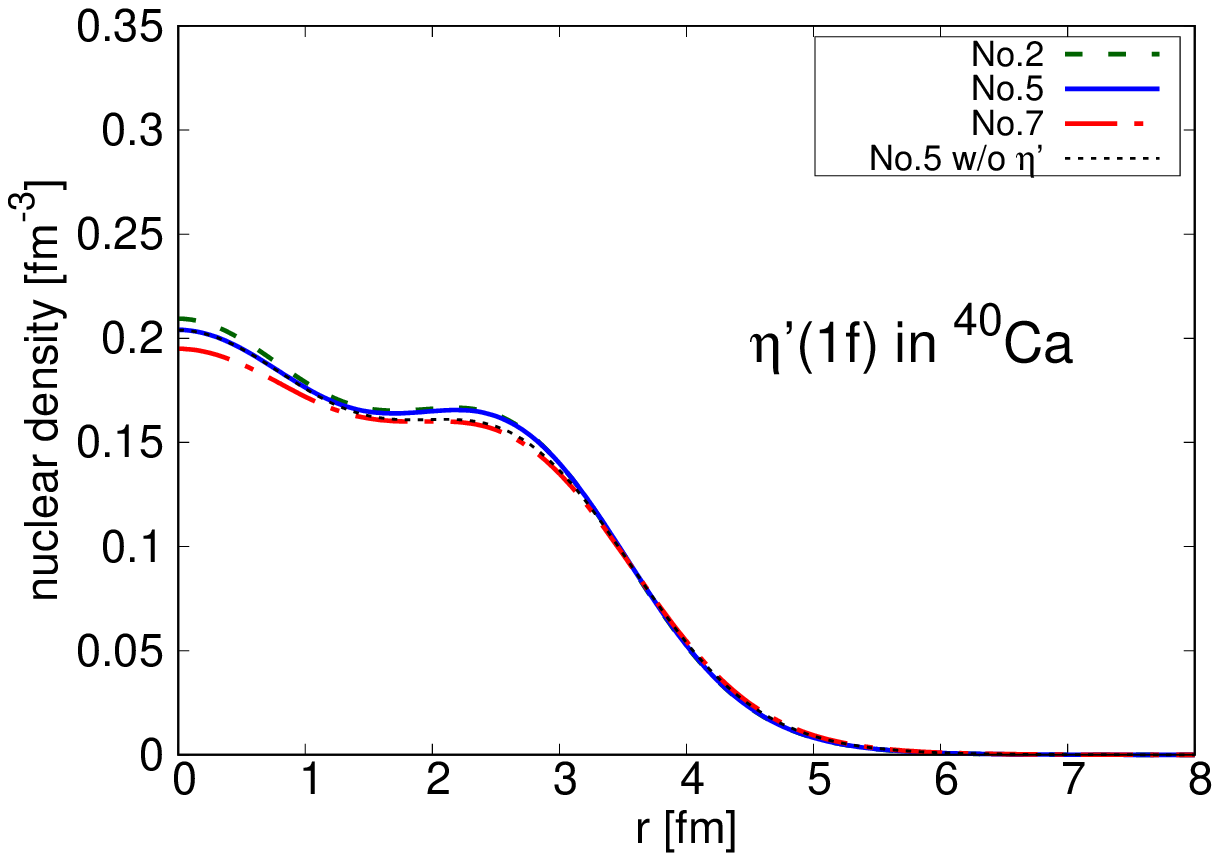}
  }
  \subfloat[1g]{
  \includegraphics[width=0.32\textwidth]{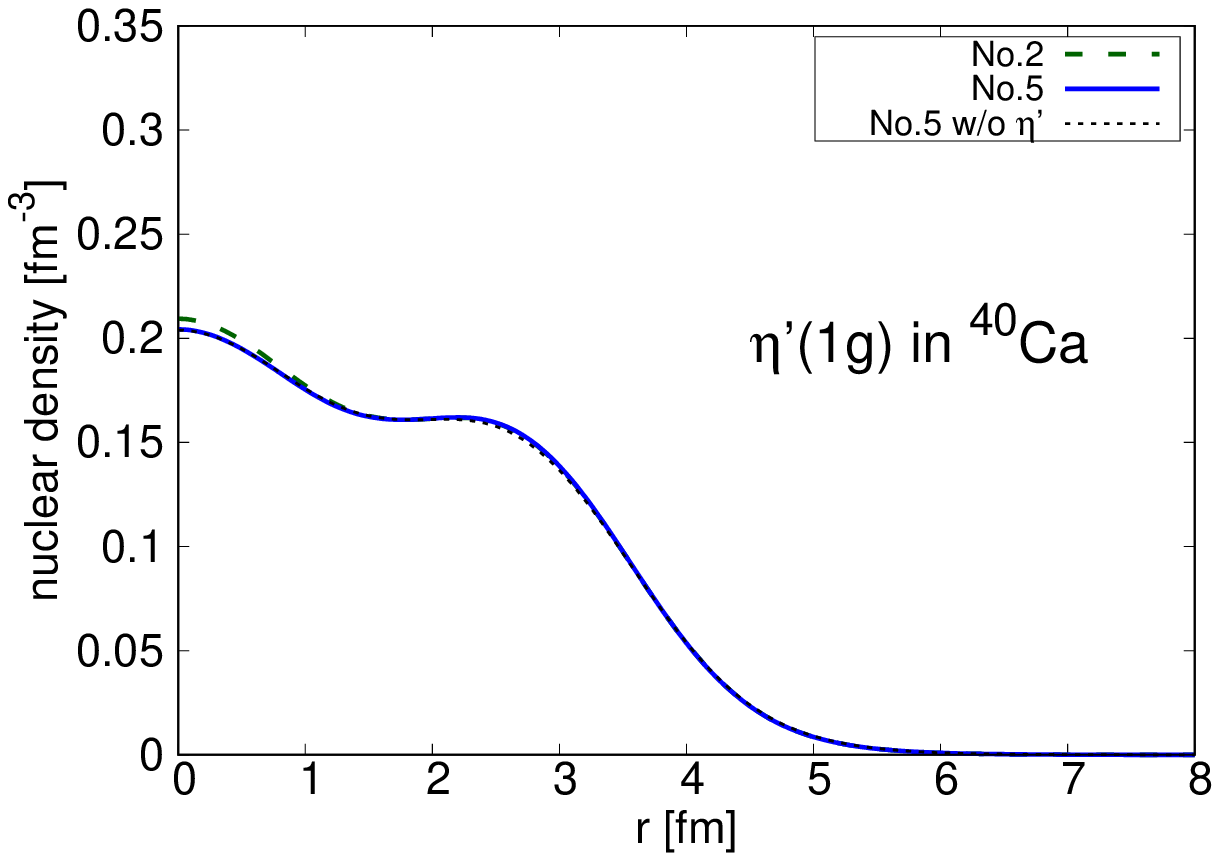}
  }
  \caption{Nuclear density profiles of the $\eta^{\prime}$ bound systems in $^{40}$Ca. The dotted 
  line stands for the density distribution of the normal nucleus $^{40}$Ca without $\eta^{\prime}$ calculated
  with parameter set 5.  \label{gr:Cadens}}
\end{figure}

Figure~\ref{gr:Caspec} shows the obtained 
binding energy spectra of the system of the $\eta^{\prime}$ meson and
the $^{40}$Ca nucleus. 
We find nine bound states for parameter set 2 and 5 and 
seven states for parameter set 7. 
We also show the details of the energy contents of the $\eta^{\prime}$-$^{40}$Ca system
in Table~\ref{tab:eta40Ca}. 
We find the similar tendency to the case of $^{16}$O. The nuclear modification effects 
are sufficiently large. With the $\eta^{\prime}$ meson, the nucleons get more binding energy.
In particular, for the case of the $\eta^{\prime}$ meson in the states
with higher angular momentum $\ell >0$, the binding energies of the $\eta^{\prime}$-nucleus system
become larger than the $\eta^{\prime}$ binding energy in magnitude. Consequently 
the energy levels of the first nodal mode get closer and might overlap each other if these states 
get a absorption width due to the strong interaction. 
In the $1s$ state larger model dependence is observed. 
In Fig.~\ref{gr:Cadens}, 
we show the nuclear density distributions for each $\eta^{\prime}$ bound state in $^{40}$Ca.
For the $1s$ bound state the central density of the $\eta^{\prime}$ mesonic nucleus
reaches to as large as 0.25 - 0.35 fm$^{-3}$. The density profiles for the other states 
are very similar to that of the normal nuclear density distribution. 

\begin{table}
\begin{center}
\caption{  \label{tab:eta40Ca}
Numerical results of the properties of the $\eta^{\prime}$-$^{40}$Ca bound system obtained 
in the present model for each $\eta^{\prime}$ bound state in units of MeV. 
The binding energy $E_{B}$ measured from the threshold of $^{40}{\rm Ca}+\eta^{\prime}$
is defined as $E_{B} = E_{\rm tot}- E_{^{40}{\rm Ca}}-m_{\eta^{\prime}}$, where the value of 
the normal $^{40}$Ca energy, $E_{^{40}{\rm Ca}}$, is taken from our calculation for the normal nucleus. 
We show here also the $\eta^{\prime}$ binding energy, $E_{\eta^{\prime}} - m_{\eta^{\prime}}$,
the average of the nucleon binding energy, $E_{\rm nuc}/A - m_{N}$, and 
the mean field energy per particle, $E_{\rm mean}/(A+1)$, where $A=40$ for $^{40}$O.
Nuclear excited stats are not considered here and nucleus stays in the ground state.  }
\begin{tabular}{l|D{.}{.}{0}D{.}{.}{0}D{.}{.}{1}|D{.}{.}{0}D{.}{.}{0}D{.}{.}{1}|D{.}{.}{0}D{.}{.}{0}D{.}{.}{1}}
\hline
 \multicolumn{1}{c|}{$\eta^{\prime}$ state} & \multicolumn{3}{c|}{$1s$}
 & \multicolumn{3}{c|}{$1p$}& \multicolumn{3}{c}{$1d$}\\
 \multicolumn{1}{c|}{model} &  \multicolumn{1}{c}{2} &   \multicolumn{1}{c}{5} &   \multicolumn{1}{c|}{7}
 &  \multicolumn{1}{c}{2} &   \multicolumn{1}{c}{5} &   \multicolumn{1}{c|}{7}
 &  \multicolumn{1}{c}{2} &   \multicolumn{1}{c}{5} &   \multicolumn{1}{c}{7} \\
 \hline
 $E_{B}$ &-77.6&-74.3&-71.7&-68.5&-67.2&-64.8&-60.0&-59.1&-56.7\\
 $E_{\eta^{\prime}} - m_{\eta^{\prime}}$ &-87.8&-79.3&-76.1&-58.8&-57.3&-54.7&-39.0&-38.3&-36.0\\
 $E_{\rm nuc}/A - m_{N}$ &-24.3&-23.6&-23.1&-24.2&-23.6&-23.1&-24.1&-23.6&-23.1\\
 $E_{\rm mean}/(A+1)$ &15.6&15.2&15.6&15.0&14.8&15.3&14.6&14.6&15.0\\
 \hline
\hline
 \multicolumn{1}{c|}{$\eta^{\prime}$ state} & \multicolumn{3}{c|}{$1f$}
 & \multicolumn{3}{c|}{$1g$}& \multicolumn{3}{c}{$2s$}\\
 \multicolumn{1}{c|}{model} &  \multicolumn{1}{c}{2} &   \multicolumn{1}{c}{5} &   \multicolumn{1}{c|}{7}
 &  \multicolumn{1}{c}{2} &   \multicolumn{1}{c}{5} &   \multicolumn{1}{c|}{7}
 &  \multicolumn{1}{c}{2} &   \multicolumn{1}{c}{5} &   \multicolumn{1}{c}{7} \\
 \hline
 $E_{B}$ &-50.7&-50.3&-47.8&-39.4&-39.3&$---$&-38.3&-37.0&-34.8\\
 $E_{\eta^{\prime}} - m_{\eta^{\prime}}$ &-19.7&-19.4&-17.5&-1.2&-1.1&$---$&-41.0&-38.9&-36.2\\
 $E_{\rm nuc}/A - m_{N}$ &-23.9&-23.5&-23.1&-23.7&-23.3&$---$&-24.1&-23.6&-23.1\\
 $E_{\rm mean}/(A+1)$ &14.2&14.3&14.7&13.8&13.9&$---$&15.2&15.1&15.5\\
 \hline
\hline
 \multicolumn{1}{c|}{$\eta^{\prime}$ state} & \multicolumn{3}{c|}{$2p$}
 & \multicolumn{3}{c|}{$2d$}& \multicolumn{3}{c}{$3s$}\\
 \multicolumn{1}{c|}{model} &  \multicolumn{1}{c}{2} &   \multicolumn{1}{c}{5} &   \multicolumn{1}{c|}{7}
 &  \multicolumn{1}{c}{2} &   \multicolumn{1}{c}{5} &   \multicolumn{1}{c|}{7}
 &  \multicolumn{1}{c}{2} &   \multicolumn{1}{c}{5} &   \multicolumn{1}{c}{7} \\
 \hline
 $E_{B}$ &-29.1&-28.3&-26.6&-15.5&-15.1&-13.4&-3.4&-3.0&-2.4\\
 $E_{\eta^{\prime}} - m_{\eta^{\prime}}$ &-17.5&-16.9&-15.5&-1.2&-1.0&-0.5&-3.4&-3.0&-2.3\\
 $E_{\rm nuc}/A - m_{N}$ &-23.8&-23.3&-22.9&-23.3&-22.9&-22.4&-23.3&-22.9&-22.4\\
 $E_{\rm mean}/(A+1)$ &14.5&14.6&15.0&14.0&14.1&14.5&14.3&14.4&14.8\\
 \hline
\end{tabular}
\end{center}
\end{table}

Finally we show the calculated results of the $\eta^{\prime}$-$^{12}$C bound system. 
Here we show the results calculated with parameter set 6, 8 and 9. In these parameter
sets, the coefficient $c$ appearing in Eq.~\eqref{eq:Lag}, which is the coefficient of the 
quadratic interaction term for the $\sigma$ field, is found to be positive. The other parameter 
sets have a negative $c$. For the normal nuclear matter the sign of the parameter $c$ is not
so important and negative values of the parameter $c$ also reproduce the normal nuclear matter properties.
Nevertheless, negative $c$ leads to unstable matter in higher densities. 
Actually, in the case of the $\eta^{\prime}$ nucleus bound system for lighter nuclei, the strong 
attraction provided by the $\eta^{\prime}$ meson is hard to be maintained by small numbers 
of nucleons and the nucleus gets unstable.  A prescription to avoid the negative value
of $c$ was proposed in Ref.~\cite{Maruyama:1993jb} by introducing a parameter.
Here we show the results obtained by the positive values of the parameter $c$. 

\begin{figure}[t]
  \centering
  \includegraphics[width=0.6\textwidth]{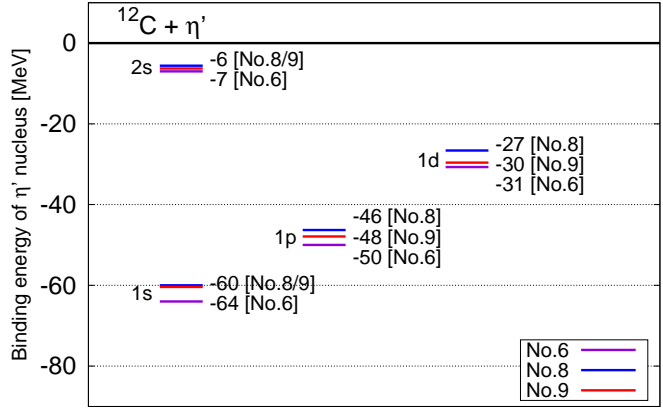}
  \caption{Binding energy spectrum of the $\eta^{\prime}$ meson in $^{12}$C. The number appearing 
  in the right side of each energy level represents the binding energy in units of MeV.
  The binding energies 
  are calculated with three parameter sets, No.6, 8 and 9. 
  In these parameter sets, the coefficient of the $\sigma^{4}$ term in the $\sigma$ self-energy
  is found to be positive. See the text for the details.  }
  \label{gr:Cspec}
\end{figure}

\begin{figure}[t]
  \centering
  \subfloat[1s]{
  \includegraphics[width=0.48\textwidth]{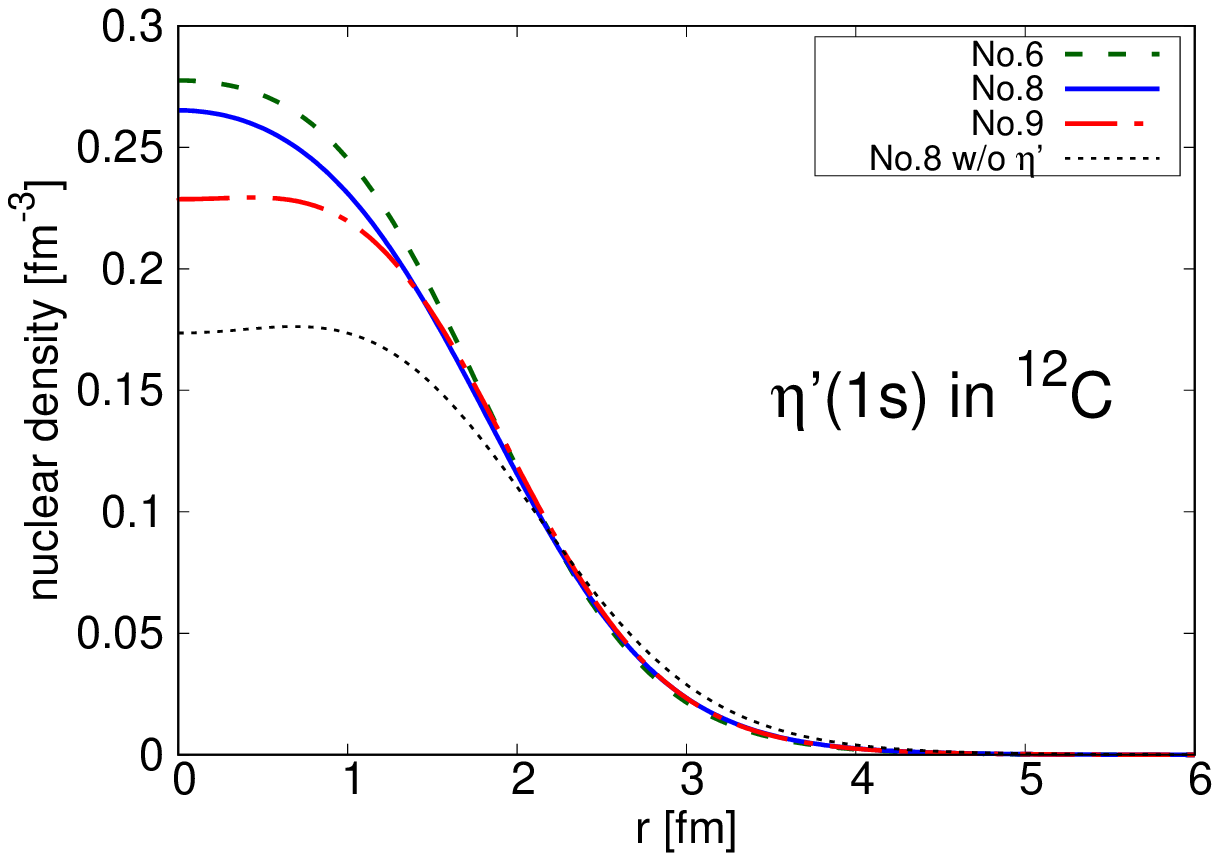}
  }
  \subfloat[1p]{
  \includegraphics[width=0.48\textwidth]{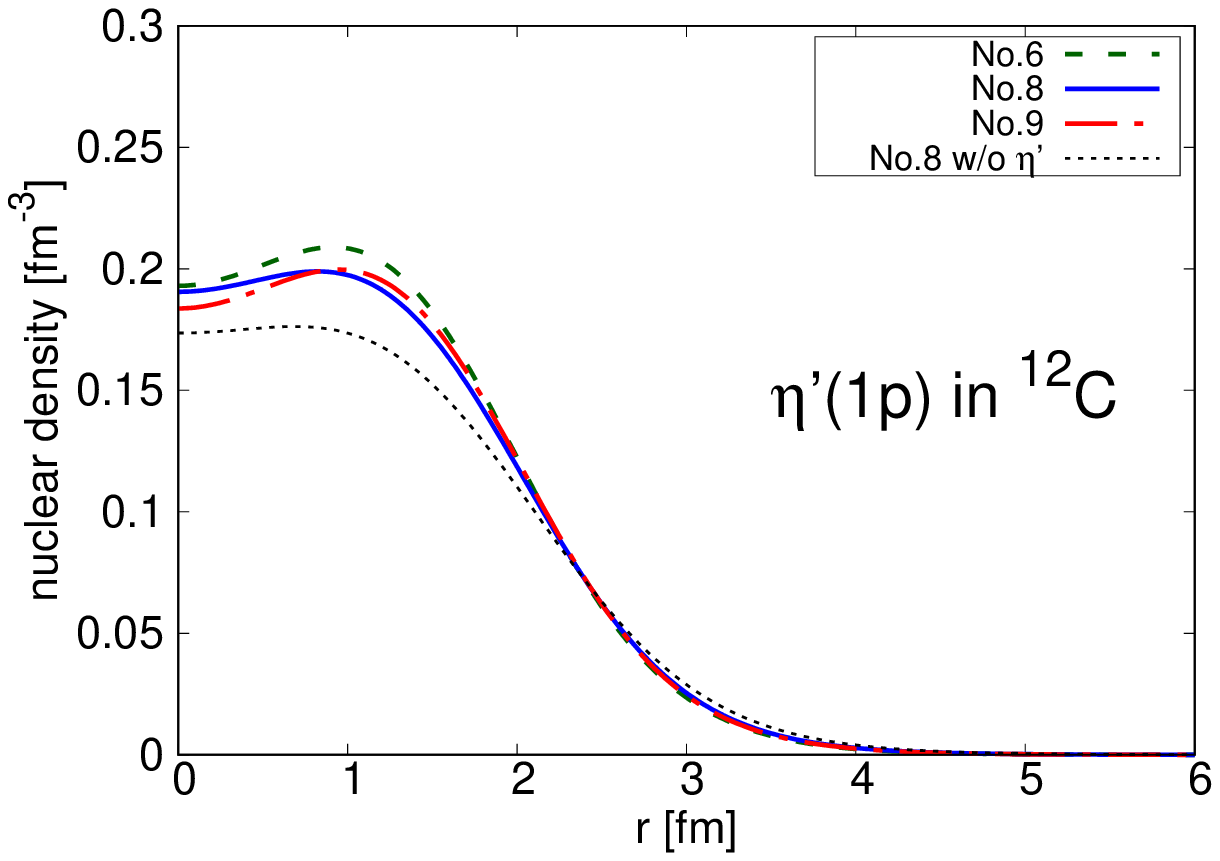}
  }\\
  \subfloat[2s]{
  \includegraphics[width=0.48\textwidth]{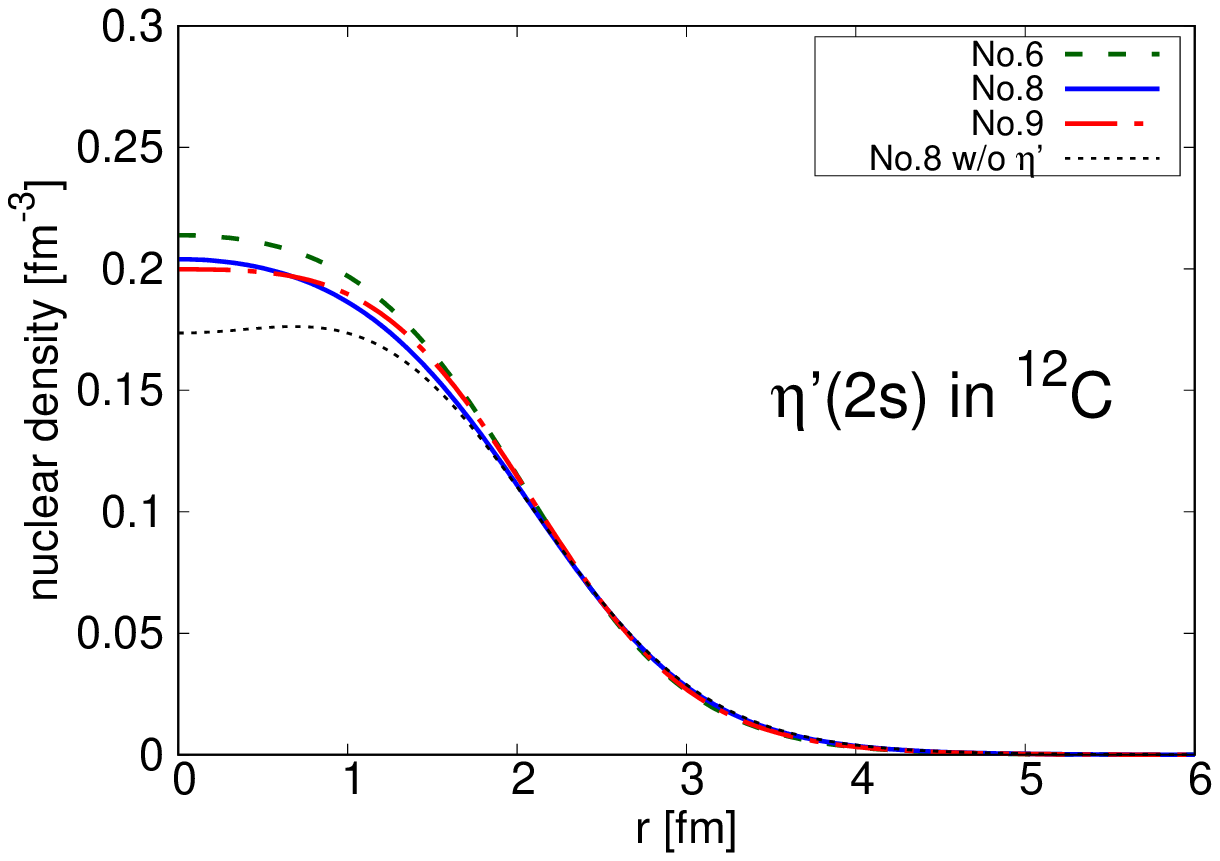}
  }
  \subfloat[1d]{
  \includegraphics[width=0.48\textwidth]{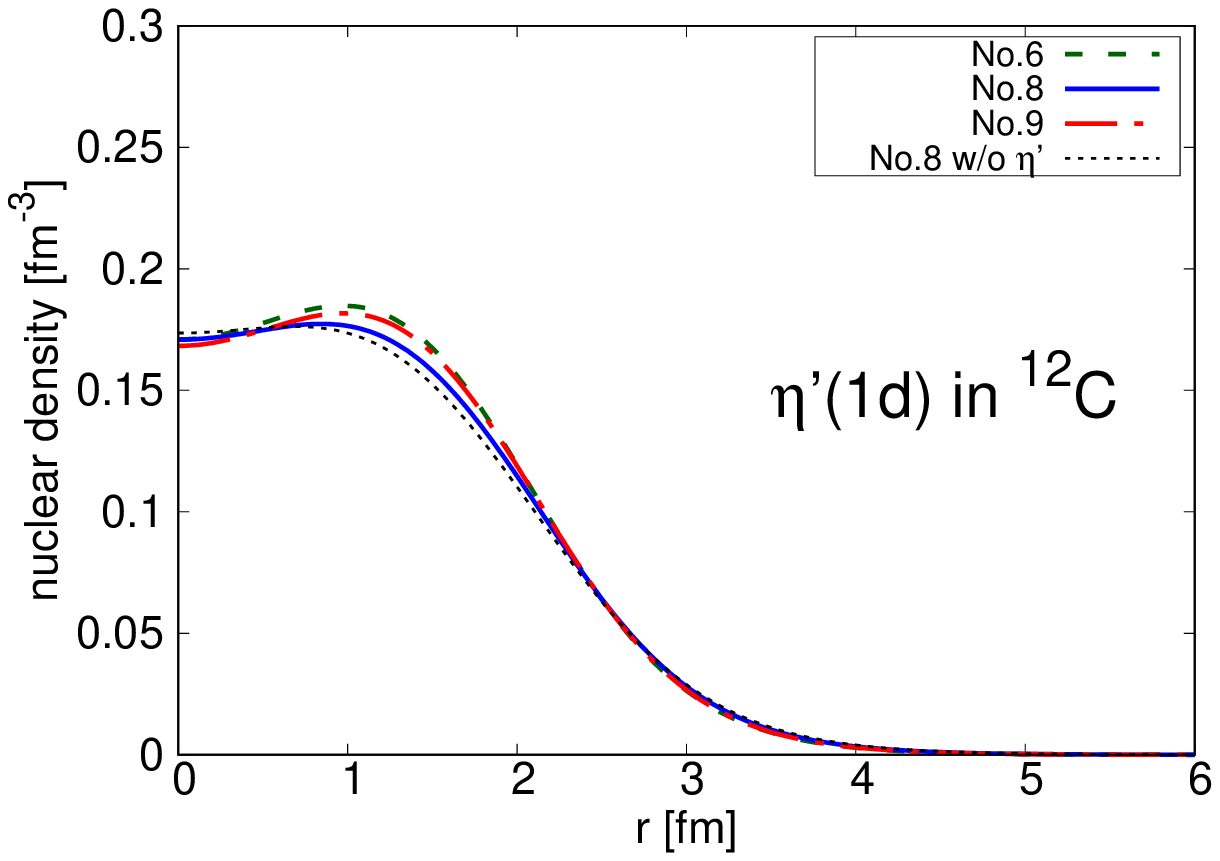}
  }
  \caption{Nuclear density profiles of the $\eta^{\prime}$ bound systems in $^{12}$C obtained with parameter set
  6, 8 and 9. The dotted
  line stands for the density distribution of the normal nucleus $^{12}$C without $\eta^{\prime}$ calculated
  with parameter set 8.  \label{gr:Cdens}}
\end{figure}

In Fig.~\ref{gr:Cspec} we show the binding energy spectrum of the $\eta^{\prime}$-$^{12}$C system
for parameter set 6, 8 and 9,
and we show the energy contents of these bound states in Table~\ref{tab:eta12C}.
There are four bound states found. 
Figure~\ref{gr:Cdens} shows the density profile of the $\eta^{\prime}$-$^{12}$C bound system.

\begin{table}
\begin{center}
\caption{  \label{tab:eta12C}
Numerical results of the properties of the $\eta^{\prime}$-$^{12}$C bound system obtained 
in the present model for each $\eta^{\prime}$ bound state in units of MeV. 
The binding energy $E_{B}$ measured from the threshold of $^{12}{\rm C}+\eta^{\prime}$
is defined as $E_{B} = E_{\rm tot}- E_{^{12}{\rm C}}-m_{\eta^{\prime}}$, where the value of 
the normal $^{12}$C energy, $E_{^{12}{\rm C}}$, is taken from our calculation for the normal nucleus. 
We show here also the $\eta^{\prime}$ binding energy, $E_{\eta^{\prime}} - m_{\eta^{\prime}}$,
the average of the nucleon binding energy, $E_{\rm nuc}/A - m_{N}$, and 
the mean field energy per particle, $E_{\rm mean}/(A+1)$, where $A=12$ for $^{12}$C.
Nuclear excited stats are not considered here and nucleus stays in the ground state.  
}
\begin{tabular}{l|D{.}{.}{1}D{.}{.}{1}D{.}{.}{1}|D{.}{.}{1}D{.}{.}{1}D{.}{.}{1}}
\hline
 \multicolumn{1}{c|}{$\eta^{\prime}$ state} & \multicolumn{3}{c|}{$1s$}
 & \multicolumn{3}{c}{$1p$}\\
 \multicolumn{1}{c|}{model} &  \multicolumn{1}{c}{6} &   \multicolumn{1}{c}{8} &   \multicolumn{1}{c|}{9}
 &  \multicolumn{1}{c}{6} &   \multicolumn{1}{c}{8} &   \multicolumn{1}{c}{9} \\
 \hline
 $E_{B}$ &-64.0&-60.0&-60.4&-50.0&-46.3&-47.9  \\
 $E_{\eta^{\prime}} - m_{\eta^{\prime}}$ &-73.6&-68.8&-67.5&-33.1&-29.8&-31.4 \\
 $E_{\rm nuc}/A - m_{N}$ &-23.9&-23.2&-22.7&-23.8&-23.0&-22.8\\
 $E_{\rm mean}/(A+1)$ & 17.4&17.8&16.8&15.2&15.6&15.1\\
 \hline
\hline
 \multicolumn{1}{c|}{$\eta^{\prime}$ state} & \multicolumn{3}{c|}{$1d$}
 & \multicolumn{3}{c}{$2s$}\\
 \multicolumn{1}{c|}{model} &  \multicolumn{1}{c}{6} &   \multicolumn{1}{c}{8} &   \multicolumn{1}{c|}{9}
 &  \multicolumn{1}{c}{6} &   \multicolumn{1}{c}{8} &   \multicolumn{1}{c}{9} \\
 \hline
 $E_{B}$ &-30.7&-26.6&-29.6&-7.0&-5.6&-6.3\\
 $E_{\eta^{\prime}} - m_{\eta^{\prime}}$ &-4.5&-2.5&-4.0&-7.8&-6.2&-7.0 \\
 $E_{\rm nuc}/A - m_{N}$ &-22.8&-21.6&-21.9&-22.1&-21.1&-21.1\\
 $E_{\rm mean}/(A+1)$ & 13.6&13.7&13.6&15.0&15.1&14.9\\
 \hline
\end{tabular}
\end{center}
\end{table}

\begin{figure}[t]
  \centering
  \includegraphics[width=\textwidth]{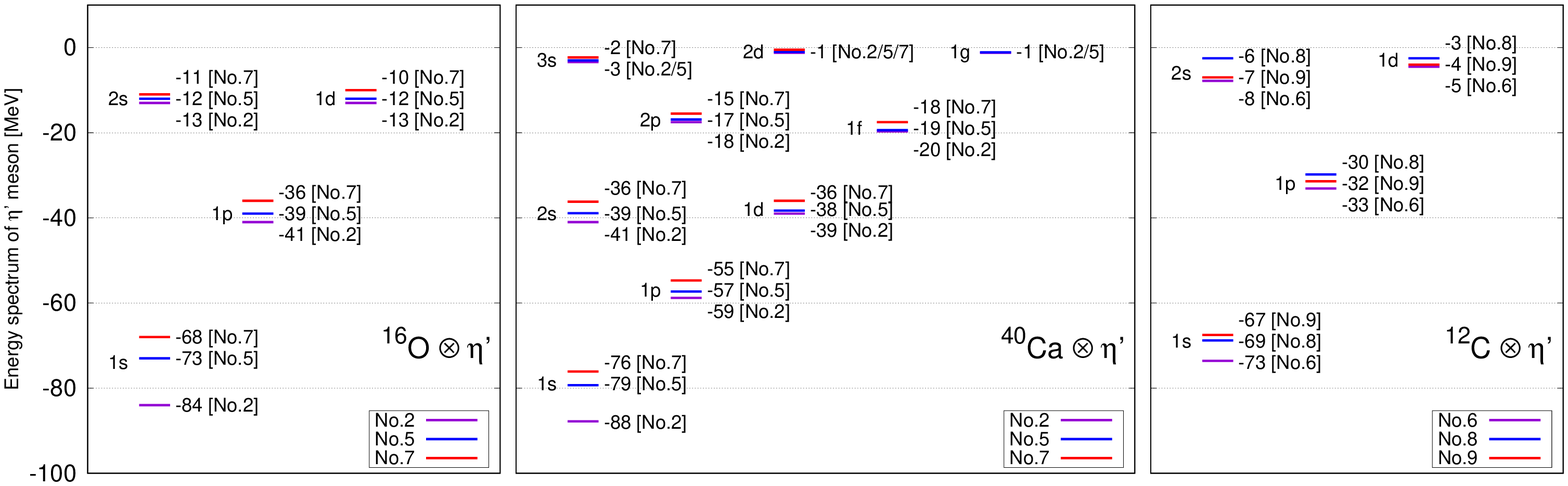}
  \caption{Binding energy spectra of the $\eta^{\prime}$ meson in $^{16}$O, $^{40}$Ca and $^{12}$C. 
  The number appearing 
  in the right side of each energy level represents the binding energy in units of MeV.
   }
  \label{gr:etaspec}
\end{figure}

It might be interesting to see the $\eta^{\prime}$ binding energy spectra themselves for the purpose 
to examine the $\eta^{\prime}$ bound state nature, even though these energies are not direct observables.
In Fig.~\ref{gr:etaspec}, we show the spectra of the $\eta^{\prime}$ binding energy, 
$E_{\eta^{\prime}}-m_{\eta^{\prime}}$, for the $^{16}$O, $^{40}$Ca and $^{12}$C nuclei.
This figure shows that the energy spectra have very similar structure to the energy spectrum of the harmonic 
oscillator potential, which is a typical pattern of finite range potentials. Because the attractive potential for the 
$\eta^{\prime}$ meson is provided by the $\sigma$ meen field,  the potential shape should be closed
to the one of the Woods-Saxon potential.   
Comparing a typical binding energy spectrum for a Woods-Saxon type potential,
in which the $2s$ and $3s$ states have less binding energies than the $1d$ and $1g$ states,
respectively,
we find comparable binding energies for $2s$ and $3s$
to those of $1d$ and $1g$, respectively. Namely the $s$ states have more attraction.
This is because for the $s$ states there are no centrifugal barrier and 
the $\eta^{\prime}$ meson can be in the center of the nucleus. There the nuclear density 
is high and the $\eta^{\prime}$ meson gets more attraction from the nuclear matter. 

\section{Conclusion}
We have investigated the $\eta^{\prime}$-nucleus bound system using the relativistic mean field approach, which is a self-contained nuclear model and provides the nuclear saturation properties within the model. Thus, the relativistic mean field theory is good for the investigation of the nuclear structure change induced by the presence of the $\eta^{\prime}$ meson inside the nucleus.
The $\eta^{\prime}$ meson mass is expected to be reduced in the nuclear matter by a scale of 100 MeV thanks to the partial (or incomplete) restoration of chiral symmetry. The mass reduction of the $\eta^{\prime}$ meson in infinite matter serves an attractive potential for the $\eta^{\prime}$ meson in finite nuclei. 
Under such hadronic scale of attractive interaction, the nuclear structure may be changed in the presence of the $\eta^{\prime}$ meson in nuclei. 
The $\eta^{\prime}$ meson is introduced into the relativistic mean field theory 
as a particle
having a coupling to the $\sigma$ field. We assume the absence of repulsive $\eta^{\prime}$-$\omega$ coupling according to the fact that there is no Weinberg-Tomozawa interaction for the $\eta^{\prime}N$ channel, which is interpreted as vector meson exchange. Owing to the attractive $\eta^{\prime}$-$\sigma$ coupling, the $\eta^{\prime}$ mass is reduced in the presence of the $\sigma$ mean field. The $\eta^{\prime}$ meson interacts with nucleons through the $\sigma$ field, and the interaction is attractive.
The strength of the $\eta^{\prime}$-$\sigma$ coupling is determined so as to reproduce the 80 MeV reduction of the $\eta^{\prime}$ mass at the saturation density. 

We have investigated the $\eta^{\prime}$ mesonic nuclei for $^{12}$C, $^{16}$O and $^{40}$Ca, and have found several bound states of the $\eta^{\prime}$ meson for each nucleus. 
The binding energy of the $\eta^{\prime}$-nucleus system is determined by not only the $\eta^{\prime}$ binding energy in the nucleus but also the change of the structure of the core nucleus. The presence of the $\eta^{\prime}$ meson 
in the nucleus attracts the $\sigma$ mean field and consequently enhances 
binding energies of nucleons in the core nucleus. 
For bound states in $s$ orbits, the $\eta^{\prime}$ meson can be at the center of the nucleus due to the absence of the centrifugal barrier. Thanks to the attractive interaction between the $\eta^{\prime}$ meson and nucleons through the $\sigma$ mean field, at the center of the nucleus, the nuclear density reaches 1.5 - 2.0 $\rho_{0}$ for  mesonic nuclei with $\eta^{\prime}$ in the $1s$ orbit depending on the nuclear matter property, soft or hard. 
For such a higher density, the repulsive mean field energy gets enhanced and the energy of the core nucleus is not so enhanced.
Because the $1s$ $\eta^{\prime}$ mesonic nucleus has such a different nuclear density configuration from the normal nucleus, it could be hard to produce experimentally the $1s$ $\eta^{\prime}$ mesonic nucleus by nuclear reactions due to small overlap of the nuclear wave functions. Along this line, it would be extremely interesting to calculate the formation cross sections by introducing the change of the nuclear structure.

\section*{Acknowledgment}

The authors thank Profs. Naoki Onishi and Tomoyuki Maruyama for telling us the details of the 
relativistic mean field theory and giving us constructive suggestions. 
D.J.\ thanks Nao Watanabe for her having initiated the study of the relativistic mean field
approach for nucleus in our group and her discussion at the preliminary stage of this work. 
The work of D.J.\ was partly supported by Grants-in-Aid for Scientific Research from JSPS (17K05449).
The work of S.H.\ was partly supported by Grants-in-Aid for Scientific Research from JSPS (16K05355).

\appendix

\section{Uniform nuclear matter}
\label{sec:NM}
In this section we explain the calculation of the nuclear matter properties by following Ref.~\cite{CompactStars}. 
We consider uniform nuclear matter, in which the mean fields have no spacial nor temporal dependences. 
We start with Lagrangian~\eqref{eq:Lag} without the photon field and the $\eta^{\prime}$ meson. 
The Euler-Lagrange equations for the mean fields are 
\begin{subequations}
\label{eq:MFeq}
\begin{eqnarray}
    m_{\sigma}^{2} \langle \sigma \rangle &=& g_{\sigma} \langle \bar \psi \psi \rangle  
    - b mg_{\sigma}^{3} \langle \sigma \rangle^{2} - c g_{\sigma}^{4} \langle \sigma \rangle^{3}, \label{eq:MFsigma}\\
    m_{\omega}^{2} \langle \omega_{0} \rangle &=& g_{\omega} \langle \psi^{\dagger} \psi \rangle, \\
    m_{\rho}^{2} \langle \rho_{0} \rangle &=& g_{\rho} \langle \psi^{\dagger} \frac{\tau^{3}}{2} \psi \rangle ,
\end{eqnarray}
\end{subequations}
where $\psi$ is the nucleon field operator, $\tau^{3}$ is the Pauli matrix for the isospin space,
and $\langle \cdots \rangle$ means the expectation 
value in the ground state of the nuclear matter. 
Hereafter we omit the brackets for the mesonic mean fields like $\sigma \equiv \langle \sigma \rangle$
for simplicity. 
The nuclear expectation values will be evaluated 
in the Fermi gas approximation later. The spacial components of the vector fields vanish in the uniform matter,
because the nuclear expectation value $\langle \bar \psi \gamma^{i} \psi \rangle$ turns to be zero.

The Dirac equation for nucleon in the presence of the mean fields reads
\begin{equation}
 \left[   ( i\delslash - g_{\omega} \omega_{0}\gamma^{0}  -  g_{\rho} \frac{\tau^{3}}{2} \rho_{0} \gamma^{0})
   - ( m - g_{\sigma} \sigma) \right] \psi(x) = 0.  \label{eq:Deqx}
\end{equation}
Because the mean fields are uniform and static, the nucleon field has a momentum eigenstate solution
\begin{equation}
   \psi(x) = \psi(k) e^{i\vec x \cdot \vec k - i \varepsilon(k) t}.
\end{equation}
Substituting this form to the Dirac equation \eqref{eq:Deqx}, we obtain
\begin{equation}
   \left[(\varepsilon(k) -V_{0})\gamma^{0} - \vec k\cdot \vec \gamma - (m-S) \right] \psi (k) = 0, \label{eq:Deqk}
\end{equation}
where we have introduced the Lorentz vector and scalar potentials
\begin{equation}
   V_{0} = g_{\omega} \omega_{0} + g_{\rho} \frac{\tau^{3}}{2} \rho_{0}, \qquad S =  g_{\sigma} \sigma, 
\end{equation}
respectively. Introducing a four-vector defined by
\begin{equation}
  K^{\mu} = (\varepsilon(k) - V_{0}, \vec k),
\end{equation}
and the effective nucleon mass as
\begin{equation}
   m^{*} = m - S =  m - g_{\sigma} \sigma,
\end{equation}
we write Dirac equation as
\begin{equation}
   (\Kslash - m^{*}) \psi(K) = 0.
\end{equation}
This corresponds to a free Dirac equation for nucleon with momentum $K$,
in which the energy and mass are shifted by the uniform vector and scalar potentials
that are provided by the mean fields, respectively. 
The solution of this Dirac equation for particle, which satisfies $K^{2} - m^{*2} = 0$,
is written as
\begin{equation}
   \psi^{(+)} (k) = \sqrt\frac{E(k) + m^{*}}{2m^{*}} 
   \left(\begin{array}{c} 1 \\ \frac{\vec \sigma \cdot \vec k}{E(k) + m^{*}} \end{array}\right)\chi_{s},
\end{equation}
which is normalized as
\begin{equation}
   \bar \psi^{(+)} \psi^{(+)} = 1,
\end{equation}
and its density is given by
\begin{equation}
   \bar \psi^{(+)} \gamma^{0} \psi^{(+)} = \frac{E(k)}{m^{*}}.
\end{equation}
Here the energy $E(k) \equiv K_{0}$ for the nucleon state with momentum $\vec k$ is given by
\begin{equation}
    E(k )  = \sqrt {\vec k^{\, 2} + m^{*2}} = \sqrt{\vec k^{\, 2} + (m - g_{\sigma} \sigma)^{2}}.
\end{equation}
It should be noted that the eigen energy of nucleon (particle) is given by
\begin{equation}
   \varepsilon(k) = E(k) + V_{0} =  \sqrt{\vec k^{\, 2} + (m - g_{\sigma} \sigma)^{2}} 
   + g_{\omega} \omega_{0} + g_{\rho} \frac{\tau^{3}}{2} \rho_{0} .
\end{equation}

In the calculation of the expectation values of nucleon in the ground state, we take Fermi gas approximation,
in which the nucleons occupies the single-particle states with $\vec k$ from the lower levels up to the Fermi 
momentum $k_{f}$. The scalar density is calculated as
\begin{eqnarray}
   \rho_{s} &=& \langle \bar \psi \psi \rangle 
   = 4  \int^{k_{f}} \frac{d\vec k}{(2\pi)^{3}} \frac{m^{*}}{E(k)} \left(\bar \psi^{(+)}(k) \psi^{(+)}(k)\right) \nonumber \\
   &=& \frac{2}{\pi^{2}} \int_{0}^{k_{f}} \frac{m-g_{\sigma}\sigma}{\sqrt {k^{2} - (m-g_{\sigma}\sigma)^{2}}} k^{2} dk,
   \label{eq:rhos}
\end{eqnarray}
where the factor 4 means the spin and isospin multiplicity and we have used the normalization of the spinor.
Here we assume the symmetric nuclear matter by setting the same Fermi momentum for proton and neutron. 
The number density of nucleon for the symmetric nuclear matter is calculated as
\begin{equation}
   \rho_{N} =\langle \psi^{\dagger} \psi \rangle 
   = 4  \int^{k_{f}} \frac{d\vec k}{(2\pi)^{3}} \frac{m^{*}}{E(k)} \left( \psi^{(+)\dagger}(k) \psi^{(+)}(k)\right) 
   = \frac{2}{\pi^{2}} \int_{0}^{k_{f}} k^{2} dk = \frac{2}{3\pi^{2}} k_{f}^{3}.
\end{equation}
With this expression, one can obtain the corresponding Fermi momentum to the nuclear density. 
For the asymmetric nuclear matter, we calculate the number densities for proton and neutron,
independently, and we obtain each Fermi momentum for proton and neutron.

The mean fields can be obtained by solving the mean field equations \eqref{eq:MFeq}, 
once the nuclear densities are given. Thus, all of the quantities are functions of the Fermi momentum, 
or the nuclear density. 
Because the scalar density $\rho_{s}$ depends on the mean field $\sigma$ through the 
effective mass of nucleon, $m^{*}$, one has to solve Eq.~\eqref{eq:MFsigma} self-consistently. 

Once one obtains the mean field configuration, one can calculate physical quantities for the nuclear matter.
The energy density is given by
\begin{eqnarray}
   {\cal E} &=& \frac12 m_{\sigma}^{2} \sigma^{2} + \frac12 m_{\omega}^{2} \omega_{0}^{2}
   + \frac12 m_{\rho}^{2} \rho_{0}^{2} - \frac13 b m g_{\sigma}^{3} \sigma^{3} 
   - \frac14 c g_{\sigma}^{4} \sigma^{4} \nonumber \\ && 
   + \frac{2}{\pi^{2}} \int_{0}^{k_{f}} \sqrt{k^{2} + m^{*2}} k^{2} d k.
\end{eqnarray}
The binding energy per nucleon is obtained by
\begin{equation}
   \frac{B}{A} = \left(\frac{\cal E}{\rho} \right) - m.
\end{equation}
We will see that $B/A$ has a minimum value at a finite density. This implies 
the saturation property of the nuclear matter. We also calculate the compressibility defined by
\begin{eqnarray}
   K = k_{f}^{2} \frac{d^{2}}{dk_{f}^{2}} \left(\frac{\cal E}{\rho} \right)
   = 9 \left. \rho^{2} \frac{d^{2}}{d\rho^{2}} \left(\frac{\cal E}{\rho} \right) \right|_{\rho=\rho_{0}}.
\end{eqnarray}
Introducing an asymmetry parameter $t = (\rho_{n}- \rho_{p})/\rho_{N}$, we calculate the symmetry 
energy as
\begin{equation}
   A_{\rm sym} = \frac12 \left. \frac{\partial^{2}}{\partial t^{2}} \left(\frac{\cal E}{\rho} \right) \right|_{t=0}.
\end{equation}

\begin{figure}[t]
\centering
\includegraphics[width=0.7\textwidth]{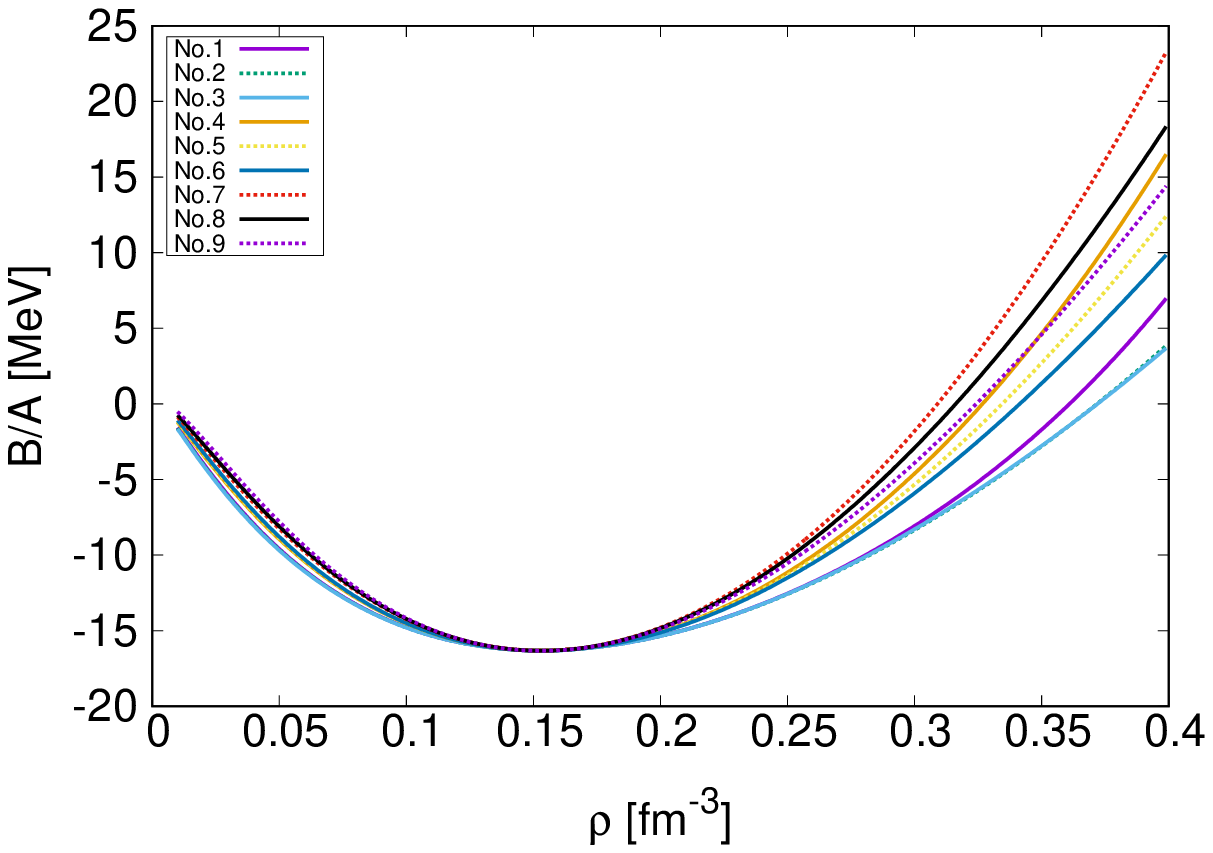}
\caption{\label{fig:saturation} Equation of state for the symmetric nuclear matter reproduced using the
parameter sets listed in Table~\ref{ta:parameters}. }
\end{figure}

The model parameters, the coupling constants $g_{\sigma}$, $g_{\omega}$, $g_{\rho}$
and the strengths of the self-interaction of the $\sigma$ field $b$ and $c$, are 
taken from Ref.~\cite{CompactStars}, in which they were 
determined so as to reproduce the nuclear matter properties, that is, the saturation density
$\rho_{0} = 0.153$~fm$^{-3}$, the binding energy per nucleon $B/A = -16.3$~MeV,
the symmetry energy $A_{\rm sym}=32.5$~MeV, the effective nucleon mass $m^{*}/m \approx 0.7$ to $0.8$,
and the compressibility $K \approx 200$ to $300$ MeV.  Nine parameter sets were proposed. 
The parameters used in this work are listed in Table~\ref{ta:parameters} and the equation 
of state for the symmetric nuclear matter reproduced with the parameters are shown in Fig.~\ref{fig:saturation}.
The meson masses itself are not relevant parameters for the nuclear matter properties,
because the change of the mass parameters in Eqs.~\eqref{eq:MFeq} can be absorbed into 
the coupling constants with their appropriate redefinition.

\section{Nuclear matter with $\eta^{\prime}$ mean field}
\label{etapNM}

In order to see the effect of the $\eta^{\prime}$ meson against the saturation properties of the nuclear
matter, we calculate equation of state of the symmetric nucleus with the $\eta^{\prime}$ mean field. 
We consider a infinite matter in which the $\eta^{\prime}$ mesons and the nucleons are uniformly 
distributed with isospin $I=0$ and the $\sigma$ and $\omega$ fields mediate their interaction.
We take the $\eta^{\prime}$ meson field as a matter field as well as the nucleon field, 
in which $\eta^{\prime}$ mesons are not annihilated nor created in the matter and are at rest. 
Because the $\eta^{\prime}$ meson is a boson, all of the $\eta^{\prime}$ particles are in the 
lowest energy state in the ground state of the matter. For the nucleon, we take Fermi gas
approximation. The matter is characterized by the 
$\eta^{\prime}$ and nuclear densities, $\rho_{\eta}$ and $\rho_{N}$, respectively. 

The equations for the $\sigma$ and $\omega$ mean fields are obtained in the same way as the
usual nuclear matter and we have 
\begin{eqnarray}
   m_{\sigma}^{2} \sigma &=& g_{\sigma} \rho_{s} + g_{\sigma\eta^{\prime}} \rho_{\eta_{\prime}}
   - b m g_{\sigma}^{3} \sigma^{2} - c g_{\sigma}^{4} \sigma^{3} \\
   m_{\omega}^{2} \omega_{0} &=& g_{\omega} \rho_{N}
\end{eqnarray}
where we have used the fact that $\eta^{\prime}$ couples only to the $\sigma$ field and 
$E_{\eta^{\prime}}= m_{\eta^{\prime}}^{*}$ in the ground state of the matter, and the nuclear scalar
density $\rho_{s}$ is calculated as Eq.~\eqref{eq:rhos} in the Fermi gas approximation. 
The effective $\eta^{\prime}$ and nucleon masses are given by
\begin{equation}
   m_{\eta^{\prime}}^{*} = \sqrt{m_{\eta^{\prime}}^{2} - 2g_{\sigma\eta^{\prime}} m_{\eta^{\prime}} \sigma}, \qquad
   m^{*} = m - g_{\sigma} \sigma,
\end{equation}
respectively.
The energy density of the matter for given densities of $\eta^{\prime}$ and nucleon is calculated as
\begin{eqnarray}
  {\cal E}(\rho_{N}, \rho_{\eta^{\prime}}) &=& 
  \frac12 m_{\sigma}^{2} \sigma^{2} + \frac12 m_{\omega}^{2} \omega_{0}^{2}
   - \frac13 b m g_{\sigma}^{3} \sigma^{3} - \frac14 c g_{\sigma}^{4} \sigma^{4} 
   \nonumber \\ && 
   + \frac{2}{\pi^{2}} \int_{0}^{k_{f}} \sqrt{k^{2} + m^{*2}} k^{2} d k + m_{\eta^{\prime}}^{*} \rho_{\eta^{\prime}}.
\end{eqnarray}
Here note that there are no $\rho$ contributions because we consider the isosinglet matter.

\begin{figure}[t]
  \centering
  \includegraphics[width=0.6\textwidth]{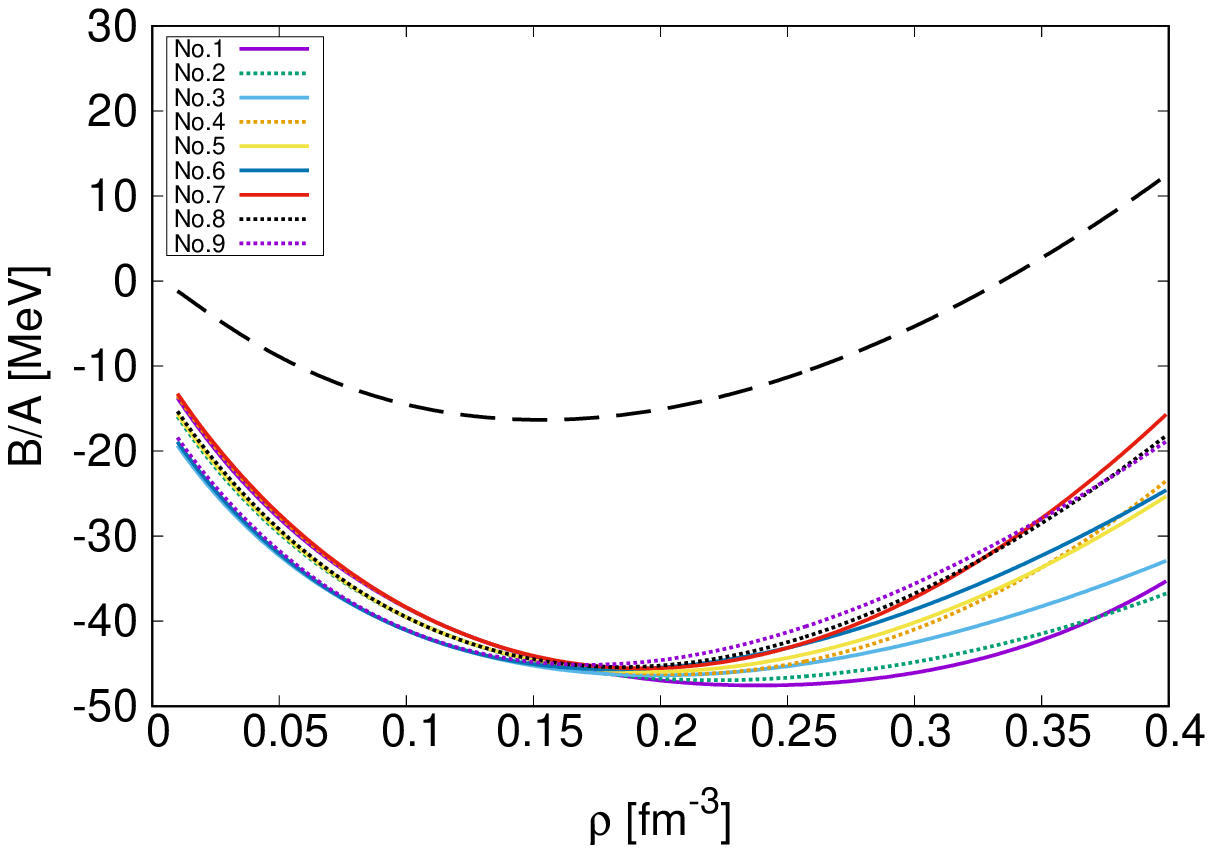}
  \caption{Equation of state for symmetric nuclear with the $\eta^\prime$ mean field as function of 
  nuclear density $\rho$. The $\eta^{\prime}$ density is fixed as $\rho_\eta^\prime=0.1$ [fm$^{-3}$]. 
  The dashed line is for the nuclear matter without $\eta^{\prime}$ calculated with parameter No.5. }
  \label{gr:gr2}
\end{figure}

Here we introduce the binding energy per particle as
\begin{equation}
   \frac{B}{A^{\prime}} = \frac{{\cal E}(\rho_{N},\rho_{\eta^{\prime}})}{\rho_{N}+ \rho_{\eta^{\prime}}}
   - \frac{\rho_{N} m + \rho_{\eta^{\prime}} m_{\eta^{\prime}}}{\rho_{N} + \rho_{\eta^{\prime}}}.
\end{equation}
Here the first term is the energy per particle and the second term is the average mass. 
In Fig.~\ref{gr:gr2}, we show equation of state for the symmetric nuclear matter with 
the $\eta^{\prime}$ meson with its density $\rho_{\eta^{\prime}} = 0.1$~fm$^{-3}$ as a function of the nuclear 
density $\rho_{N}$. 
The density $\rho_{\eta^{\prime}} = 0.1$ fm$^{-3}$ corresponds to the $\eta^{\prime}$ density 
at the center of the $\eta^{\prime}$ mesonic
nucleus. The dashed line is the binding energy per nucleon for the usual nuclear matter without $\eta^{\prime}$,
which is calculated with parameter 5, for comparison. 
This figure shows that we find the saturation property in the matter at
higher density  with larger binding energy than the normal nuclear matter.
In Table~\ref{ta:ta2} we show the values of the saturation densities $\rho_{0}$, 
the binding energy per particle $B/A^{\prime}$, the compressibility $K$, the symmetry energy $A_{\rm sym}$,
the nucleon effective mass $m^{*}$ and the $\eta^{\prime}$ effective mass $m_{\eta^{\prime}}^{*}$
for each parameter set. 
Thanks to the presence of the $\eta^{\prime}$ matter, equation of state is changed from the usual nuclear matter.
In particular, the saturation density gets higher and the compressibility gets larger. 

\begin{table}[t]
\begin{center}
\caption{\label{ta:ta2} Properties of nuclear matter with $\eta^{\prime}$ mean field at saturation density $\rho_{0}$. 
The $\eta^{\prime}$ density is fixed as $\rho_\eta^\prime=0.1$ [fm].
We show the saturation density $\rho_{0}$, the binding energy per particle $B/A^{\prime}$, 
the compressibility $K$, the symmetry energy $A_{\rm sym}$, and the effective masses of nucleon and $\eta^{\prime}$, $m^{*}$ and $m_{\eta^{\prime}}^{*}$.}
 \begin{tabular}{c|ccccccccc}\hline
No.& 1  &2  &3  &4  &5  &6  &7  &8  &9  \\
\hline
$\rho_0$ [fm$^{-3}$] & 0.239  &0.221  &0.198  &0.207  &0.196  &0.181  &0.193  &0.184  &0.171  \\
$B/A^{\prime}$ [MeV]&  -47.56  &-46.95  &-46.37  &-46.24  &-45.93  &-45.66  &-45.55  &-45.39  &-45.13  \\
$K$ [MeV] & 445.6  &390.4  &361.5  &527.7  &463.7  &400.8  &565.5  &497.2  &422.2  \\
$A_{\rm sym}$ [MeV] & 37.5  &32.8  &28.0  &30.5  &27.9  &24.9  &27.7  &25.6  &23.1  \\
${m^*}/{m}$& 0.496  &0.610  &0.714  &0.562  &0.651  &0.734  &0.590  &0.670  &0.747  \\
$m^{*}_{\eta^{\prime}}/m_{\eta^{\prime}}$& 0.855 &0.866 &0.878 &0.876 &0.881 &0.887 &0.884 &0.888 &0.893 \\
	\hline
	\end{tabular}
\end{center}
\end{table}


%

\vfill\pagebreak

\end{document}